\documentclass[structabstract]{aa}
\usepackage{graphicx}
\usepackage{txfonts}
\usepackage{longtable,lscape}

\usepackage{amssymb}
\usepackage{natbib}

\def\msun{$M_\odot$}
\newcommand{\reac}[6]{$\rm\,{}^{#1}\kern-0.8pt{#2}\,({#3}\,,{#4})\, {}^{#5}\kern-0.8pt{#6}\,$}
\newcommand{\chem}[2]{$\mathrm{^{#1}#2}$}
\defcitealias{costa2006}{Paper II}
\defcitealias{pumo2006}{Paper I}

\begin{document}
   \title{Convective overshooting and production of s-nuclei in massive stars during their core He-burning phase\thanks{Tables  \ref{tab_overproduc_M} and \ref{tab_overproduc_Z} are only available in electronic form at the CDS via anonymous ftp to cdsarc.u-strasbg.fr (130.79.128.5) or via http://cdsweb.u-strasbg.fr/cgi-bin/qcat?J/A+A/.}}
   \titlerunning{Convective overshooting and production of s-nuclei}
   \author{M.L. Pumo\inst{1,2} \and G. Contino\inst{3} \and A. Bonanno\inst{2} \and R.A. Zappal\`a\inst{3}}
   \institute{INAF - Osservatorio Astronomico di Padova, Vicolo dell'Osservatorio 5, I-35122 Padova, Italy\\
              \and INAF - Osservatorio Astrofisico di Catania, Via S. Sofia 78, I-95123 Catania, Italy\\
              \and Universit\`a di Catania, Dip. di Fisica e Astronomia (Sez. Astrofisica), Via S. Sofia 78, I-95123 Catania, Italy\\
              }
   \offprints{M.L. Pumo, \email{mlpumo@oact.inaf.it}}
   \date{Received XXXX XX, XXXX; accepted XXXX XX, XXXX}

  \abstract
   {Many studies on the s-process and, more specifically, on the s-process weak component have been performed so far, but a detailed scrutiny of the impact of the stellar evolution modeling uncertainties on the efficiency of this nucleosynthesis process is still missing.}
   {We analyze the role of convective overshooting on the production of s-nuclei in massive stars during their core He-burning phase.}
   {With the ``post-processing'' technique we explore the role of the convective overshooting on the production of s-nuclei in stellar models of different initial mass and metallicity ($15 \leq M_{ZAMS}/M_{\odot} \leq 25$; $10^{-4} \leq Z \leq 0.02$), considering a range of values for the parameter $f$, which determines the overall efficiency of convective overshooting.}
   {We find enhancements in the production of s-nuclei until a factor $\sim 6$ (measured as the average overproduction factor of the 6 s-only nuclear species with $60\lesssim A\lesssim90$) in all our models of different initial mass and metallicity with $f$ in the range $0.01\mbox{-}0.035$ (i.e. models with overshooting) compared to the production obtained with ``no-overshooting'' models (i.e. models with the same initial mass and metallicity, but $f=10^{-5}$). Moreover the results indicate that the link between the overshooting parameter $f$ and the s-process efficiency is essentially monotonic in all our models of different initial mass and metallicity. Also evident is the higher s-process efficiency when we progressively increase for a given f value both the mass of the models from 15 M$_\odot$ to 25 M$_\odot$ and the Z value from 10$^{-4}$ to 0.02. We also briefly discuss the possible consequences of these results for some open questions linked to the s-process weak component efficiency, as well as a ``rule of thumb'' to evaluate the impact of the convective overshooting on the yields of a generation of stars.}
   {}

   \keywords{Nuclear reactions, nucleosynthesis, abundances -- Convection -- Stars: evolution -- Stars: interiors}

   \maketitle

\section{Introduction}
\label{sec_intro}

As first pointed out by \citet{b2fh}, it is now widely accepted that about half of the elements between Fe and Bi are formed via the so-called s-process, through neutron capture reactions and beta decays along the ``valley of stability''. It is also known that more than one s-process ``component'' (i.e. an event with a single set of physical conditions like neutron exposure, initial abundances and neutron density) is required in order to explain the observed solar distribution of s-nuclei abundances.

Current views on the subject suggest the existence of two components (so-called main and weak component of s-process, respectively) that, in terms of stellar environments, correspond to two distinct categories of stars in different evolutionary phases \citep[e.g.][]{kappeler99}. In particular, the main component is associated with low-mass stars ($M_{ZAMS} \sim 1.5-3 M_\odot$) during their asymptotic giant branch (AGB) phase, and the main neutron source is the $^{13}$C($\alpha$,n)$^{16}$O reaction; while the other one occurs in massive stars ($M_{ZAMS} \gtrsim 13 M_{\odot}$) primarily during their core He-burning phase, and the main neutron source is the $^{22}$Ne($\alpha$,n)$^{25}$Mg reaction.

In addition to these two components, other kinds of stars, such as massive AGB ($M_{ZAMS} \sim 4-7 M_\odot$) and super-AGB stars ending their life as NeO white dwarfs ($M_{ZAMS} \sim 7.5-10 M_\odot$; for details see e.g. Fig. 1 in \citealt{pumo2009b}, but also \citealt{ps07} or \citealt{pumo07} and references therein), could also contribute to the nucleosynthesis of s-species, but this hypothesis still needs further investigation \citep[][and references therein]{pumo2009a}. Moreover, in some studies \citep*[][]{gallino98,busso99,lugaro03,gorielysiess2004} the existence of a ``strong'' component is also suggested, which occurrs in low-metallicity stars of low-intermediate mass during the AGB phase, which is supposed to be responsible for the synthesis of ``massive'' (around \chem{208}{Pb}) s-species. Furthermore, \citet[][]{travaglio04} propose the existence of an additional component referred to as lighter element primary s-process (LEPP), but its nature is still unclear and under debate \citep*[e.g.][and references therein]{tur09,pignatari10}.

Concerning the weak component, there is a wide consensus about the main characteristics of this nucleosynthesis process and, in particular, about its sensitivity to stellar mass and metallicity \citep[e.g.][]{kappeler99}. As for the dependence on the stellar mass, quantitative studies (see e.g. \citealt*{prantzos90}; \citealt{kappeler94}; \citealt{rayethashimoto2000}; \citealt{the2000}, \citeyear{the07}) show that the s-process weak component efficiency decreases with decreasing initial stellar mass, and that the shape of the distribution of the overproduction factors as a function of the mass number essentially does not depend on the initial stellar mass value. This behavior is connected to the fact that the reaction \reac{22}{Ne}{\alpha}{n}{25}{Mg} becomes efficient only for $T\gtrsim 2,5\times 10^8~K$, so the production of s-nuclei is more and more efficient when the initial stellar mass is increased, because more massive models burn helium at a ``time averaged'' higher temperature; however the ratio of the overproduction factor F$_i$ of a given s-only nucleus $i$ to the average overproduction factor F$_0$ (see Section \ref{sec_results} for details on F$_i$ and F$_0$) remains fairly constant irrespective of the stellar mass, so the shape of the distribution of the overproduction factors does not change when the initial stellar mass is increased. As for the effect of metallicity, the s-process weak component efficiency depends on the so-called {\it source/seed} ratio \footnote{Considering that the \chem{22}{Ne} is the main neutron provider during the core He-burning phase and neglecting all the heavier \chem{56}{Fe} nuclei, one obtains: $$ \mbox{source/seed} \simeq \mbox{\chem{22}{Ne}/\chem{56}{Fe}}.$$ This last quantity approximately corresponds to the \chem{14}{N}/\chem{56}{Fe} ratio at the end of the core H-burning phase which, in turn, is roughly equal to the O/\chem{56}{Fe} ratio at the ZAMS \citep[see][for details]{prantzos90,rayethashimoto2000}.} \citep[see e.g.][]{prantzos90,rayethashimoto2000}. If the {\it source/seed} ratio is constant with the metallicity Z, the efficiency is expected to increase when increasing the Z value, because the effect of the \chem{16}{O} primary poison becomes less important when the abundances of the {\it source} nuclei increase with Z. For a non-constant {\it source/seed} ratio that increases when decreasing Z, the efficiency (measured in terms of the number of neutrons captured per initial \chem{56}{Fe} seed nucleus n$_c$; see also Section \ref{sec_results}) has a non-linear behavior with Z, which reflects the interplay between two opposite factors: from one hand, the aforementioned role of the \chem{16}{O} primary poison, which tends to decrease n$_c$ with decreasing Z, because its abundance remains the same independently of Z, so its relative importance increases as Z decreases; on the other hand the effect of the increased {\it source/seed} ratio, which tends to increase n$_c$ with decreasing Z, because the number of available neutrons per nucleus {\it seed} increases as Z decreases.

Although the general features of the s-process weak component seem to be  well established therefore, there are still some open questions linked to the nuclear physics, the stellar evolution modeling, and the possible contribution to the s-nucleosynthesis from post-He-burning stellar evolutionary phases (see e.g. \citealt*{woosley2002}; \citealt{pumo2006}, hereafter Paper I; \citealt{costa2006}, hereafter Paper II).

Many works \citep*[see e.g.][]{MA93,kappeler94,rayethashimoto2000,the2000,the07,hoffman2001,tur07,tur09,pignatari10,bennett10} have been devoted to analyze the uncertainties due to nuclear physics, linked both with the reaction rates of reactions affecting the stellar structure evolution (as, for example, the triple-alpha, the \reac{12}{C}{\alpha}{\gamma}{16}{O} and the \chem{12}{C} + \chem{12}{C} reactions) and with reaction rates on which the so-called ``neutron economy'' (i.e. the balance between neutron emission and captures) is based. The contribution to the synthesis of s-nuclei during the post-core-He-burning evolutionary phases was also explored by many authors (see e.g. \citealt*{arcoragi1991}; \citealt{raiteri1993}; \citealt{the2000}, \citeyear{the07}; \citealt{hoffman2001}; \citealt{raucher02}; \citealt{LC03}; \citealt{tur07}, \citeyear{tur09}), including in some cases the explosive burning.

As for the impact of uncertainties owing to stellar evolution modeling, the determination of the size of the convective core and, more in general, of the mixing regions represents one of the major problems in calculating stellar structure and its evolution \citep[see e.g.][]{zahn91,young03}, which may directly affect the efficiency of the s-process nucleosynthesis by influencing the chemical and local temperature stratification \citep[see e.g.][]{canuto97,MF94,DX08}, by determining the amount of stellar material which experiences neutron irradiation \citep[see e.g.][]{langer89}, and by giving rise to a variation of the s-process lifetime \footnote{The s-process lifetime essentially represents the time duration where the physical conditions are suitable to synthesize s-nuclei and, more in general, to change their abundance. Following the definition adopted in \citetalias{pumo2006} and \citetalias{costa2006}, the s-process lifetime can be evaluated according to the relation $$ s-
\mbox{process lifetime} = t_{fin\_spro} - t_{ini\_spro},$$ with $t_{ini\_spro}$ defined as the time in which the temperature at the stellar center T$_c$ is logT$_c$=8, and $t_{fin\_spro}$ defined as the time where the extension of the convective core is reduced to zero and logT$_c$=8.6.} (see e.g. \citetalias{costa2006}). The convective core's extension of a star with a given initial mass and metallicity is determined in turn by a series of physical parameters such as the choice of the convective instability criterion (Schwarzschild's or Ledoux's criteria), the extra mixing processes induced by axial rotation and convective overshooting (see e.g. \citealt*{Chiosi92}; \citealt{woosley2002}).

A series of studies have been devoted to analyze the effects of these physical parameters on the evolution of massive stars (see e.g.  \citealt{MM97}, \citeyear{MM00}; \citealt*{Heger00}; \citealt{woosley2002}; \citealt*{Hirschi04}; \citealt{LC06}; \citealt*{eleid09}) and to examine the corresponding impact on the s-process weak component (see e.g. \citealt{langer89}; \citetalias{pumo2006}; \citetalias{costa2006}; \citealt{pignatari08}). As far as the convective overshooting is concerned, one finds that this extra mixing process leads to an increase of the convective core mass and to a variation of the chemical and temperature stratification that in turn tend to enhance the s-process weak component efficiency by giving rise to an increase of the amount of material that experiences neutron irradiation and to a variation of the s-process lifetime (see e.g. \citetalias{costa2006} for more details); however, a detailed scrutiny of the role of convective overshooting on the production of s-nuclei in massive stars is still missing.

In the light of our preparatory studies on this topic (\citetalias{pumo2006} and \citetalias{costa2006}), which show a not negligible impact of the convective overshooting on the s-process during core He-burning in a $25$ M$_\odot$ star (ZAMS mass) with an initial metallicity of $Z=0.02$, we believe it is worthwhile examining this issue further by analyzing the s-process efficiency in stellar models of different initial mass and metallicity ($15 \leq M_{ZAMS}/M_{\odot} \leq 25$; $10^{-4} \leq Z \leq 0.02$).

We are aware that a full understanding of how the convective overshooting affects the s-nucleosynthesis process efficiency requires studies on the stellar evolution (possibly including explosive burning), fully describing the correlated nucleosynthesis processes. Nevertheless, we believe that the first step is an analysis of the impact of the convective overshooting on the s-process during core He-burning, which represents the primary evolutionary phase where the physical conditions are suitable for the neutron capture nucleosynthesis in massive stars.

To perform this analysis, as we did already in \citetalias{pumo2006} and \citetalias{costa2006}, we use a diffusive approach to model the convective overshooting. According to this approach \citep*[for details see e.g.][and reference therein]{freytag96,herwing97}, a free parameter --- the so-called overhooting parameter $f$ --- determines the efficiency of the extra mixing due to the convective overshooting, so that a higher value of $f$ implies a bigger extension of the extra mixing outside the convective region (see also Sect. \ref{sect_code}). \citet*[][]{salasnich99} give some indications on how to fix the parameter $f$ in stellar models of massive stars and, in particular, claim a setting of this parameter to a value of $\sim 0.015$ in order to reproduce the observed distribution of massive stars across the HR diagram; however, other works \citep*[see e.g.][]{young01,DX08} indicate that massive starts could have a more extended overshooting region (compared to what is obtained using $f=0.015$), that the $f$ value may be mass dependent and, consequently, that a higher $f$ value cannot be excluded for massive stars. Thus, since the proper $f$ value is still widely debated, we prefer to consider a whole range of its possible values ($f$\footnote{In some previous studies about the s-process weak component \citep[e.g.][]{arcoragi1991}, the approach used to model the convective overshooting is based on an artificial enhancement of the convective core extension by a quantity $\alpha_{ov} \times H_p$, where $\alpha_{ov}$ is a free parameter and $H_p$ is the pressure scale height estimated at the upper radial edge of the convective core established through the Schwarzschild criterion. The extent of the overshooting region obtained with this instantaneous mixing approach corresponds approximately to what is expected using a diffusive approach when setting $f=0.1 \times \alpha_{ov}$ \citep[see e.g.][]{salasnich99,herwing97}.}$= 0.01$, $0.02$ and $0.035$ for models with overshooting, and $f=10^{-5}$ for models without overshooting).

\section[]{Input physics}
\subsection{The stellar evolution code}
\label{sect_code}

The stellar data were calculated starting from ZAMS until the end of core He-burning with the same stellar evolution code described in \citetalias{pumo2006} and \citetalias{costa2006} \citep*[see also][for details]{Weiss00,bonanno02}, but with the \reac{12}{C}{\alpha}{\gamma}{16}{O} reaction rate taken from NACRE (Nuclear Astrophysics Compilation of REaction rates, \citealt{angulo1999}). A complete description of the input physics used for calculating the stellar models can be found in \citetalias{costa2006}; however we recall the main features concerning the treatment of the mixing here.

The mixing is treated as a diffusive process, and accordingly nuclear species abundance changes are calculated with the equation
\begin{equation}
\label{diffusion}
\frac{dX}{dt} = \left(\frac{\partial X}{\partial t}\right)_{nuc} + \frac{\partial}{\partial m_r}
\left[\left(4\pi r^2 \rho\right)^2 D \frac{\partial X}{\partial m_r} \right]_{mix},
\end{equation}
where the first term on the right is the time derivative of a given isotopic abundance (mass fraction) owing to nuclear reactions, while the second is the diffusive term that describes mixing.

The difference among convective, overshooting, and radiative regions lies in the relation used to evaluate the value of the diffusion coefficient $D$. In convective zones (established through the Schwarzschild criterion) $D$ is given by
\begin{equation}
D_{conv} = \frac{1}{3} v_c l,
\end{equation}
where $v_c$ is the average velocity of convective elements derived according to the mixing length theory and $l=\alpha \times H_p$ is the mixing length ($H_p$ is the pressure scale height and $\alpha$ is mixing length parameter put to 1.7 in our calculations). Beyond convective zones (overshooting regions), the following relation is used instead:
\begin{equation}
\label{diffusion_coefficient_2}
D_{over} = D_0 exp \frac{-2z}{H_v} \,\,\,\,\,\,\, \mbox{with} \,\,\,H_v = f\cdot H_p,
\end{equation}
where $D_0$ is the value of $D_{conv}$ at the upper radial edge of the convective core, $z = |r - r_{edge}|$ is the radial distance from the same edge, and $f$ is the so-called overshooting parameter, which determines the overall efficiency of convective overshooting. For $z \gg 1$ (radiative regions) the diffusion coefficient is $\sim 0$, and abundance changes are only due to the nuclear reaction term in Eq. (\ref{diffusion}).

\subsection{The nucleosynthesis code}

The s-nucleosynthesis code, the s-process network, and the coupling of nucleosynthesis simulations with stellar evolution data are the same as described in \citetalias{pumo2006} and \citetalias{costa2006} in detail. However, concerning the s-process network, we recall that it includes 472 nuclides (up to \chem{210}{Po}) linked by 834 reactions, and allows us to follow the nucleosynthesis of all the s-species up to \chem{209}{Bi}. Moreover, as for the coupling of nucleosynthesis simulations with stellar evolution data, we recall that the ``post-processing'' technique is used according to the prescriptions described by \citet*{prantzos87} in addition to those reported in \citet{kippenan} to take into account the inclusion of not previously mixed material into the ``mixing '' zones (convective plus overshooting regions) when these latter expand.

\section{Models}

We performed 22 s-process simulations considering two grids of stellar models (see also Table \ref{tab_model}) with different initial (i.e. at ZAMS) masses ($15 \leq M_{ZAMS}/M_{\odot} \leq 25$), initial metallicities\footnote{The initial mass fractions of metals for the Z=0.02 models are taken to be equal to the values used in Papers I and II to make our results comparable to those reported in these papers. The initial mass fractions of metals for the models with Z$< 0.02$ are scaled from the values of the Z=0.02 models in a way that their relative abundances are the same as in the Z=0.02 models and, consequently, are given by the relation $$ X_i(Z)=X_i(Z=0.02)\times\frac{Z}{0.02},$$ where $X_i(Z)$ is the initial mass fraction of $ith$ metal for the models with Z$< 0.02$ and $X_i(Z=0.02)$ is the initial mass fraction of the same element for the Z=0.02 models.} ($10^{-4} \leq Z \leq 0.02$) and overshooting parameter values ($f$=10$^{-5}$, 0.01, 0.02 and 0.035).

\begin{table}
  \centering
  \caption{Stellar models}
  \begin{tabular}{@{}llcll@{}}
  \hline\hline
  Grid & Set & $M_{ZAMS}\,[M_\odot]$ & $Z$    & $f$      \\
  \hline
    (1)& (a) &15                     &0.02     & 10$^{-5}$\\
       &     &15                     &0.02     & 0.01     \\
       &     &15                     &0.02     & 0.02     \\
       &     &15                     &0.02     & 0.035    \\
       & (b) &20                     &0.02     & 10$^{-5}$\\
       &     &20                     &0.02     & 0.01     \\
       &     &20                     &0.02     & 0.02     \\
       &     &20                     &0.02     & 0.035    \\
       & (c) &25                     &0.02     & 10$^{-5}$\\
       &     &25                     &0.02     & 0.01     \\
 \hline
    (2)& (a) &20                     &10$^{-4}$& 10$^{-5}$\\
       &     &20                     &10$^{-4}$& 0.01     \\
       &     &20                     &10$^{-4}$& 0.02     \\
       &     &20                     &10$^{-4}$& 0.035    \\
       & (b) &20                     &0.005    & 10$^{-5}$\\
       &     &20                     &0.005    & 0.01     \\
       &     &20                     &0.005    & 0.02     \\
       &     &20                     &0.005    & 0.035    \\
       & (c) &20                     &0.01     & 10$^{-5}$\\
       &     &20                     &0.01     & 0.01     \\
       &     &20                     &0.01     & 0.02     \\
       &     &20                     &0.01     & 0.035    \\
 \hline
\end{tabular}
\label{tab_model}
\end{table}

The first grid, composed of models with a given initial metallicity (see set (a), (b), and (c) of grid (1) in Table \ref{tab_model}), permits us to devote particular attention to the impact of convective overshooting in stellar models of different initial mass. Since a detailed study on the impact of the convective overshooting on the s-process of a $Z$=$0.02$, $M$=$25\, M_{\odot}$ stellar model was already performed in \citetalias{pumo2006} and \citetalias{costa2006}, here we repeated the s-process simulations for this stellar model considering only two value of $f$ (see set (c) of the grid (1) in Table \ref{tab_model}), in order to study the effect of the change of the \reac{12}{C}{\alpha}{\gamma}{16}{O} reaction rate in the stellar evolution code.

The second grid, which is composed of models with a given initial mass (see set (a), (b), and (c) of grid (2) in Table \ref{tab_model}), coupled with the set (b) of grid (1) in Table \ref{tab_model}, gives us the opportunity to analyze the role of the core overshooting in stellar models of different initial metallicity.

\section{Parameters describing the s-process efficiency and results}
\label{sec_results}

The s-process efficiency was analyzed in terms of the same s-process efficiency indicators used in \citetalias{pumo2006} and \citetalias{costa2006}, namely:

\begin{itemize}
 \item[-] the average overproduction factor F$_0$ for the 6 s-only nuclei \chem{70}{Ge}, \chem{76}{Se}, \chem{80}{Kr}, \chem{82}{Kr}, \chem{86}{Sr} and \chem{87}{Sr}, given by $$ F_0=\frac{1}{N_s}\sum_{i}F_i \mbox{\,~with\,~ } F_i=\frac{X_i}{X_{i,ini}},\,\,N_s=6$$ where F$_i$ is the overproduction factor, X$_i$ is the mass fraction (averaged over the convective He-burning core) of s-only nucleus $i$ at the end of s-process, X$_{i,ini}$ is the initial mass fraction of the same nucleus, and N$_s$ is the number of the s-only nuclei within the mass range $60 \leq A \leq 87$;
 \item[-] the maximum mass number A$_{max}$ for which the species in the $60 \leq A \leq A_{max}$ mass range are overproduced by at least a factor of about 10 and 5 (first and second value respectively in the third column of Tables \ref{tab_z0.02} and \ref{tab_m20});
 \item[-] the number of neutrons captured per initial \chem{56}{Fe} seed nucleus n$_c$;
 \item[-] the maximum convection zone mass extension (hereafter MCZME) during the core He-burning s-process;
 \item[-] the duration of core He-burning s-process.
\end{itemize}

\begin{table*}
  \centering
  \caption{Parameters describing the s-process efficiency (see text) for the first grid of stellar models (grid (1) in Table \ref{tab_model}).}
 \begin{tabular}{llrrcccrr}
  \hline\hline
      &$f$       &$F_0$     &$A_{max}$ & \multicolumn{3}{c}{$n_c$} &$MCZME$&$Duration$ [$sec$] \\   
      &          &          &          & This work & K94   & T07    &       &                   \\
  \hline
  (a) &$10^{-5}$ &$9.80 $   &$87-88$   &$1.19$     &$1.80$ &$1.19$  &$1.89M_{\odot}$&$5.44\times10^{13}$ \\
      &$0.01$    &$15.45$   &$88-90$   &$1.80$     &$....$ &$....$  &$2.54M_{\odot}$&$5.42\times10^{13}$ \\
      &$0.02$    &$27.32$   &$88-90$   &$2.50$     &$....$ &$....$  &$2.90M_{\odot}$&$5.16\times10^{13}$ \\
      &$0.035$   &$55.96$   &$88-94$   &$3.35$     &$....$ &$....$  &$3.56M_{\odot}$&$4.40\times10^{13}$ \\
 \hline                                                            
  (b) &$10^{-5}$ &$43.85 $  &$88-94 $  &$3.03$     &$3.66$ &$2.34$  &$3.26M_{\odot}$&$3.77\times10^{13}$ \\
      &$0.01$    &$49.31 $  &$89-94 $  &$3.19$     &$....$ &$....$  &$3.94M_{\odot}$&$3.73\times10^{13}$ \\
      &$0.02$    &$90.10 $  &$91-96 $  &$3.90$     &$....$ &$....$  &$4.41M_{\odot}$&$3.65\times10^{13}$ \\
      &$0.035$   &$172.56$  &$92-100$  &$4.74$     &$....$ &$....$  &$4.81M_{\odot}$&$3.59\times10^{13}$ \\
 \hline
  (c) &$10^{-5}$ &$92.92 $  &$89-94 $  &$3.96$     &$5.41$ &$3.52$  &$5.40M_{\odot}$&$2.32\times10^{13}$ \\
      &$0.01$    &$164.72$  &$92-100$  &$4.68$     &$....$ &$....$  &$6.48M_{\odot}$&$2.13\times10^{13}$ \\
 \hline
\end{tabular}
\label{tab_z0.02}
\tablefoot{Sets (a), (b), and (c) refer to the Z=0.02 stellar models with M$_{ZAMS}=15~M_\odot$, $20~M_\odot$, and $25~M_\odot$, respectively. The overshooting parameter value used in the stellar evolution code is reported in the first column for each set of stellar models with a fixed initial mass. The values of $n_c$ obtained by \citet{kappeler94} and \citet{the07} for their Z=0.02 stellar models with a similar mass and input physics are also reported for comparison (see columns labeled as K94 and T07, respectively). The sign ``....'' means datum not available.}
\end{table*}

\begin{table*}
  \centering
  \caption{As in Table \ref{tab_z0.02}, but for the second grid of stellar models (grid (2) in Table \ref{tab_model}).}
  \begin{tabular}{llrrcccrr}
  \hline\hline
      &$f$        &$F_0$     &$A_{max}$ &$n_c$   &$MCZME$          &$Duration$ [$sec$]  \\
  \hline
  (a) &$10^{-5}$  &$3.78 $   &$86-87$   &$0.16$  &$3.52M_{\odot}$  &$3.58\times10^{13}$  \\
      &$0.01$     &$4.30 $   &$86-87$   &$0.23$  &$4.54M_{\odot}$  &$3.57\times10^{13}$  \\
      &$0.02$     &$4.27 $   &$86-87$   &$0.23$  &$5.19M_{\odot}$  &$3.18\times10^{13}$  \\
      &$0.035$    &$4.34 $   &$86-87$   &$0.24$  &$5.58M_{\odot}$  &$3.08\times10^{13}$  \\
 \hline
  (b) &$10^{-5}$  &$9.56  $  &$87-88 $  &$1.19$  &$3.96M_{\odot}$  &$4.04\times10^{13}$  \\
      &$0.01$     &$10.47 $  &$88-89 $  &$1.31$  &$4.46M_{\odot}$  &$3.78\times10^{13}$  \\
      &$0.02$     &$11.06 $  &$88-89 $  &$1.40$  &$4.93M_{\odot}$  &$3.21\times10^{13}$  \\
      &$0.035$    &$11.14 $  &$88-89 $  &$1.40$  &$5.26M_{\odot}$  &$2.89\times10^{13}$  \\
 \hline
  (c) &$10^{-5}$  &$48.05 $  &$89-92 $  &$3.08$  &$3.94M_{\odot}$  &$4.09\times10^{13}$  \\
      &$0.01$     &$72.14 $  &$89-94 $  &$3.65$  &$4.15M_{\odot}$  &$3.79\times10^{13}$  \\
      &$0.02$     &$68.27 $  &$91-94 $  &$3.57$  &$4.80M_{\odot}$  &$3.40\times10^{13}$  \\
      &$0.035$    &$125.44$  &$92-98 $  &$4.29$  &$5.23M_{\odot}$  &$3.14\times10^{13}$  \\
 \hline
\end{tabular}
\label{tab_m20}
\tablefoot{Sets (a), (b), and (c) refer to the M=20$M_\odot$ stellar models with Z=10$^{-4}$, 0.005, and 0.01, respectively. The overshooting parameter value used in the stellar evolution code is reported in the first column for each set of stellar models with a fixed initial metallicity.}
\end{table*}

\begin{figure}
 \includegraphics[width=85mm]{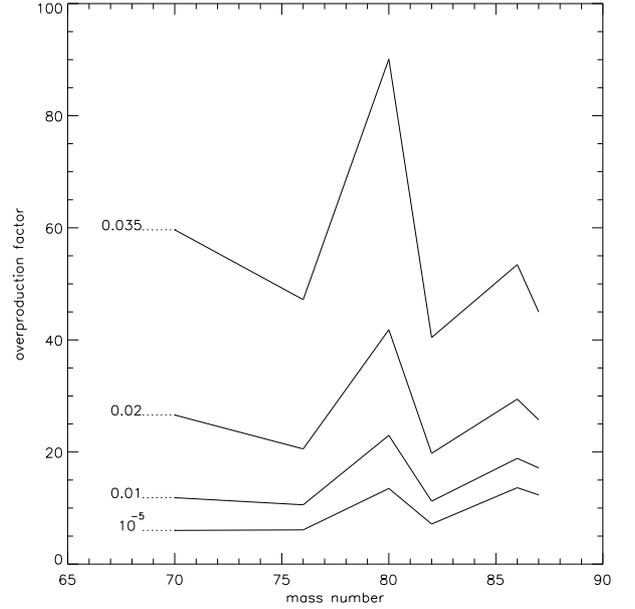}
 \caption{Overproduction factor for the six s-species \chem{70}{Ge}, \chem{76}{Se}, \chem{80}{Kr}, \chem{82}{Kr}, \chem{86}{Sr}, and \chem{87}{Sr}, for Z=0.02, M=15$M_\odot$ stellar models with overshooting parameter $f$=10$^{-5}$, 0.01, 0.02, and 0.035 (see labels).}
\label{fig_m15}
\end{figure}

\begin{figure}
 \includegraphics[width=85mm]{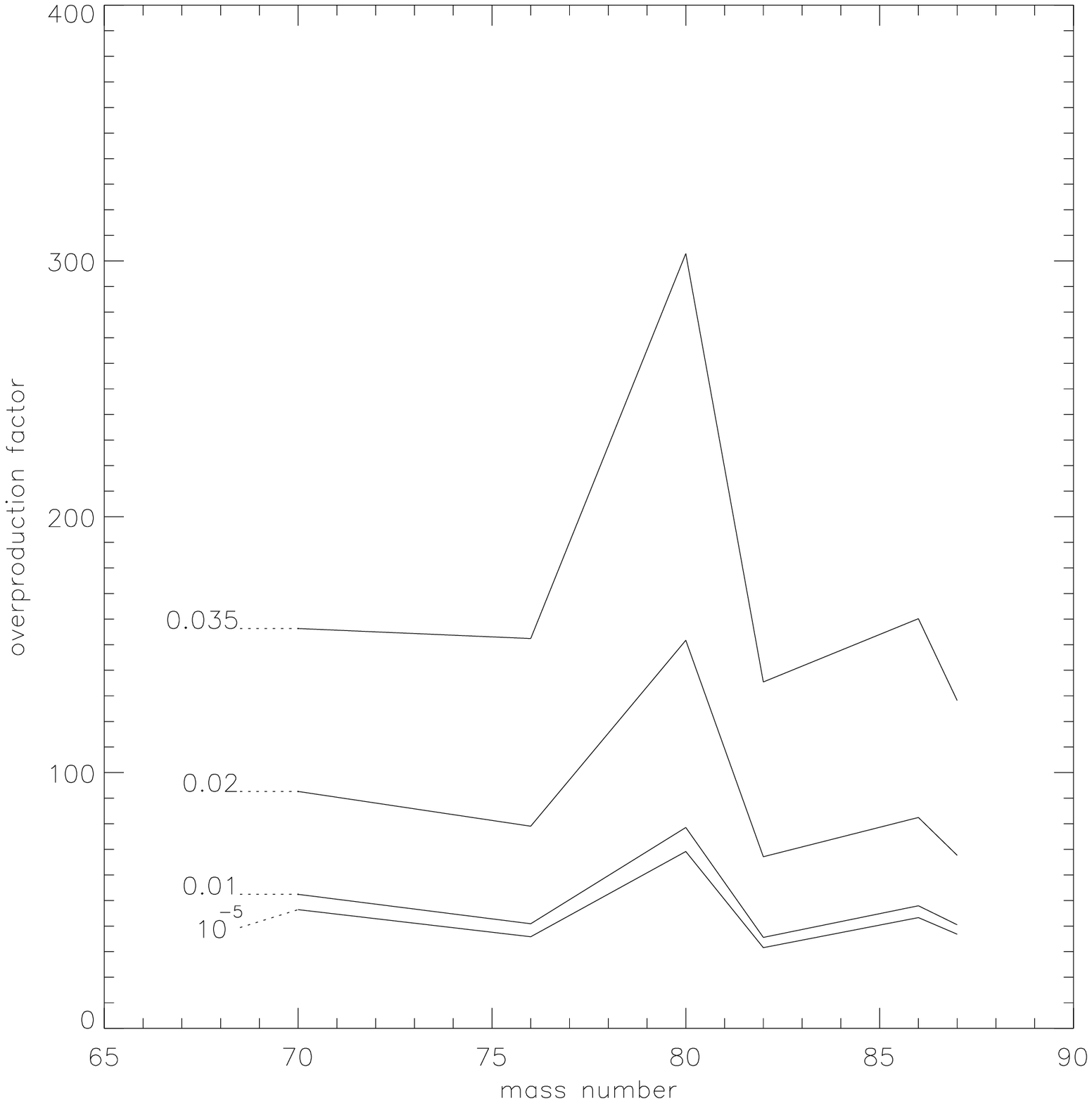}
 \caption{As in Fig. \ref{fig_m15}, but for Z=0.02, M=20 $M_\odot$ stellar models.} 
\label{fig_m20}
\end{figure}

\begin{figure}
 \includegraphics[width=85mm]{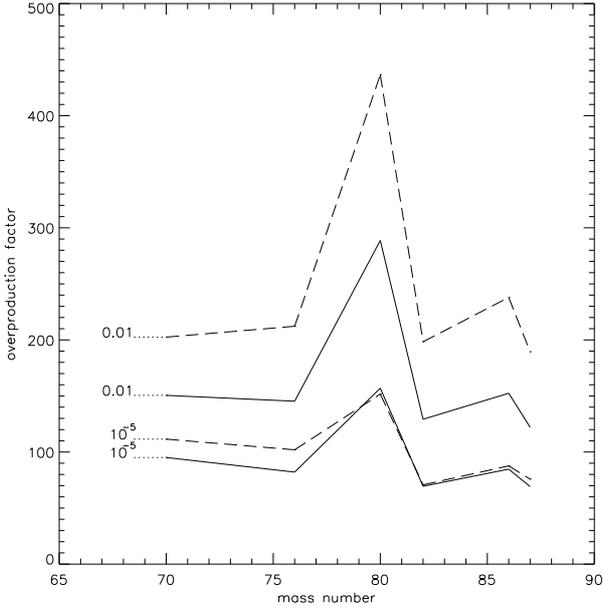}
 \caption{As in Fig. \ref{fig_m15}, but for Z=0.02, M=25 $M_\odot$ stellar models with overshooting parameter $f$=10$^{-5}$ and 0.01 (see labels). Solid lines shown our results obtained with stellar models using the NACRE rate for \reac{12}{C}{\alpha}{\gamma}{16}{O}; while dashed lines are data from \citetalias{pumo2006} and \citetalias{costa2006}, using an older rate.} 
\label{fig_m25}
\end{figure}

\begin{figure}
 \includegraphics[width=85mm]{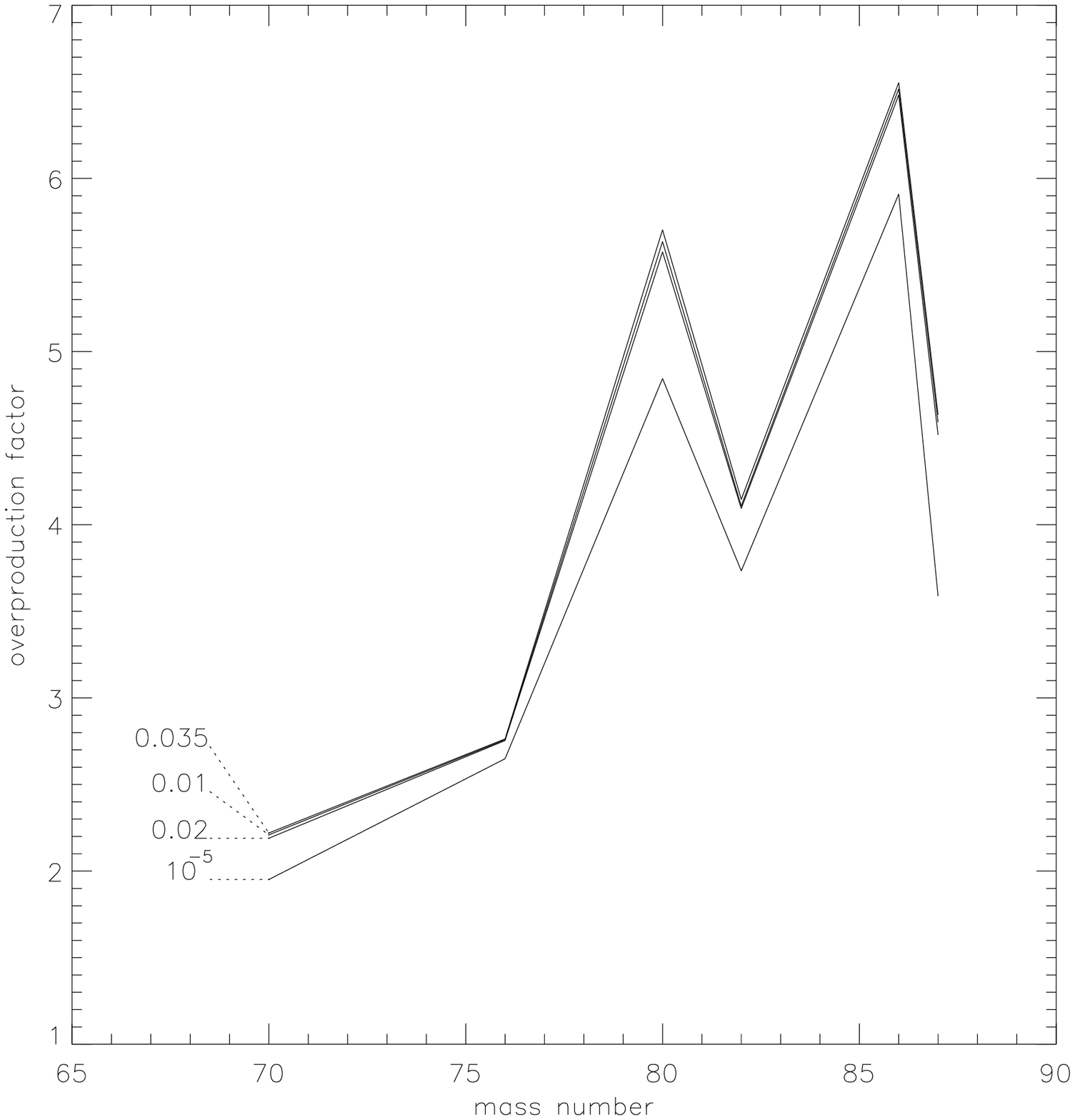}
 \caption{As in Fig. \ref{fig_m15}, but for Z=10$^{-4}$, M=20$M_\odot$ stellar models.}
\label{fig_z10m4}
\end{figure}

\begin{figure}
 \includegraphics[width=85mm]{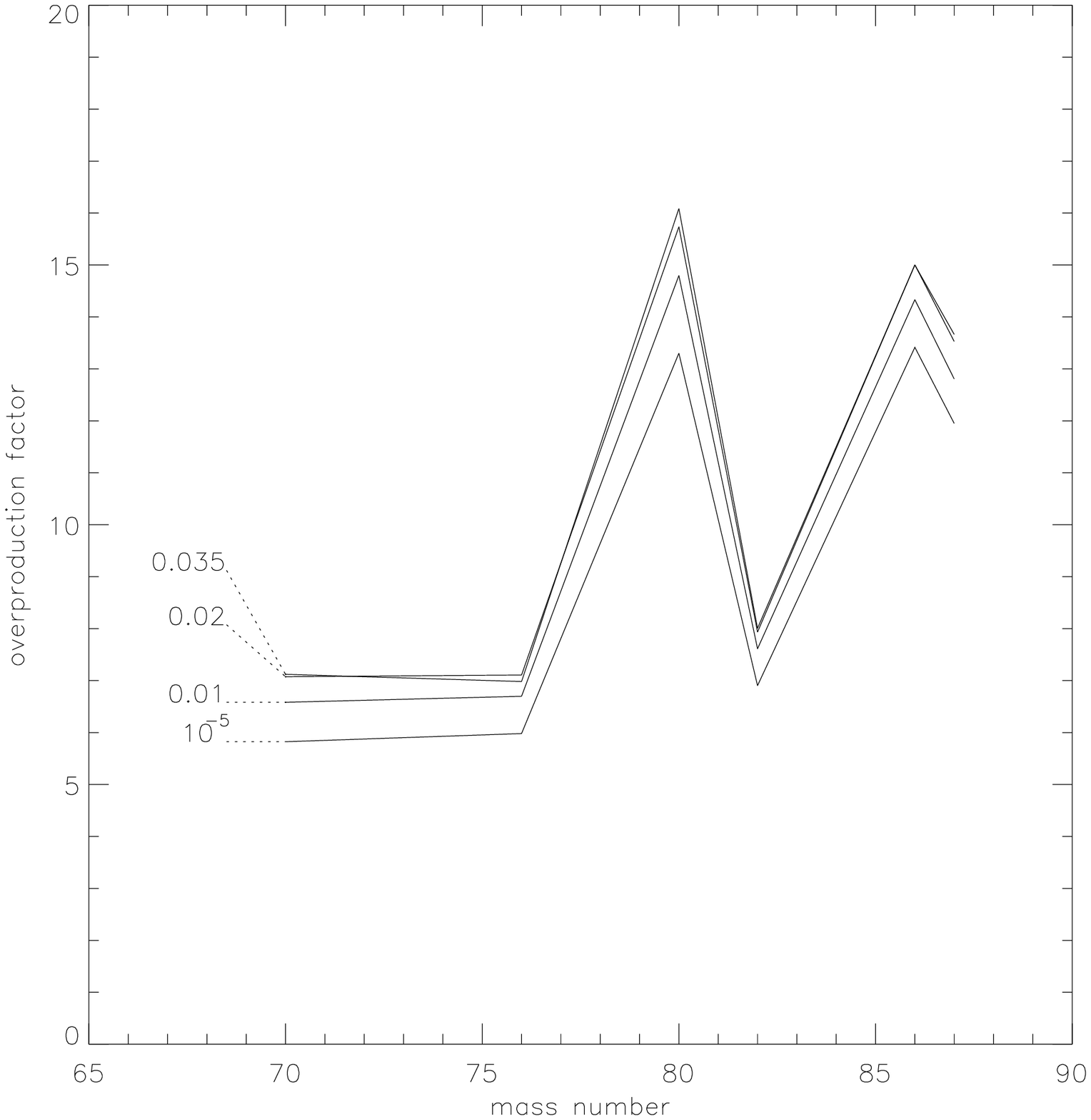}
 \caption{As in Fig. \ref{fig_m15}, but for Z=0.005, M=20 $M_\odot$ stellar models.} 
\label{fig_z0.005}
\end{figure}

\begin{figure}
 \includegraphics[width=85mm]{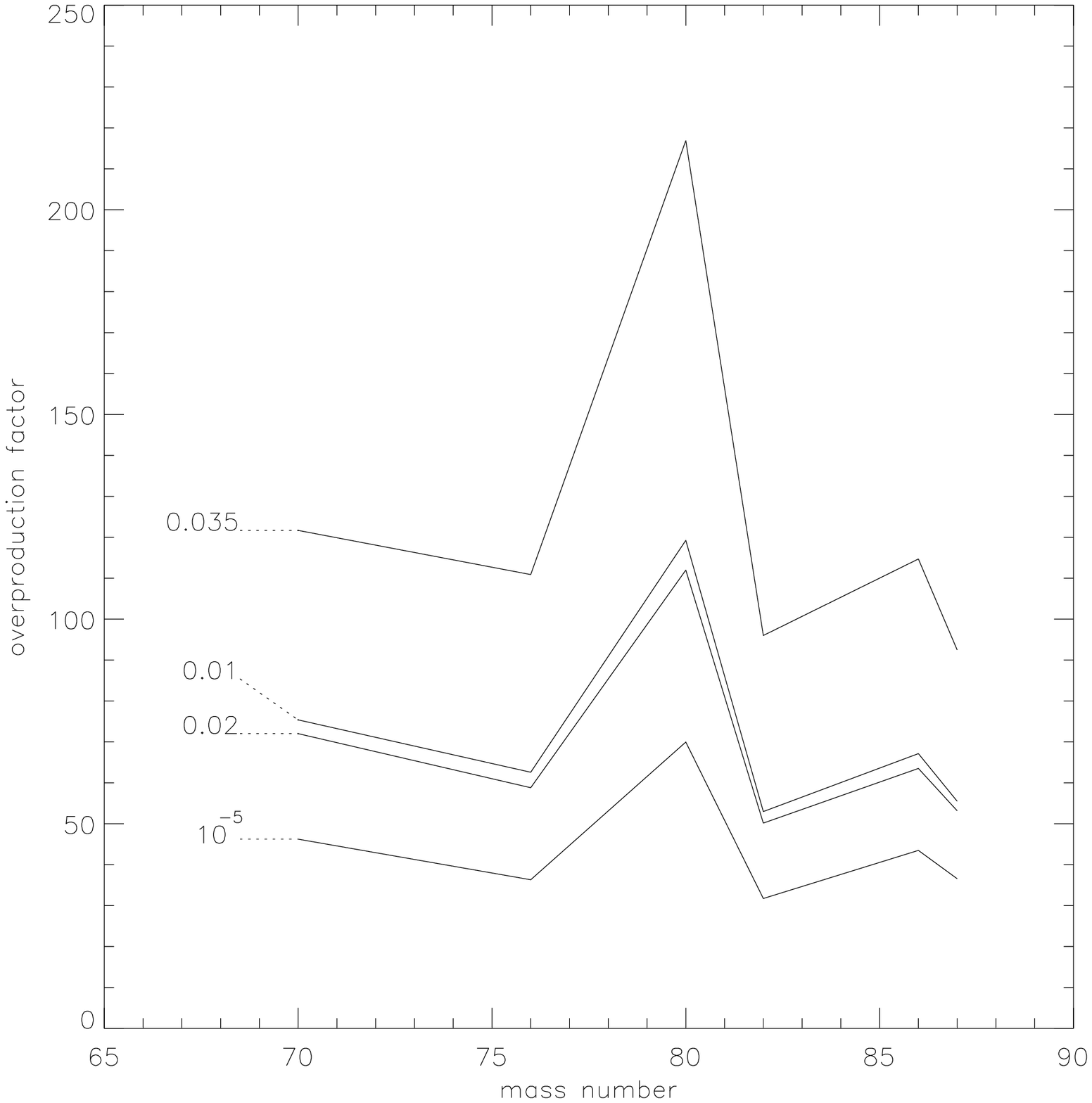}
 \caption{As in Fig. \ref{fig_m15}, but for Z=0.01, M=20 $M_\odot$ stellar models.} 
\label{fig_z0.01}
\end{figure}

The results for the first grid (i.e. grid (1) in Table \ref{tab_model}) are summarized in terms of the previous parameters in Table \ref{tab_z0.02}, and the overproduction factors of the s-only nuclei within the mass range $60 \leq A \leq 87$ as a function of nuclear mass number A are reported in Figures \ref{fig_m15} to \ref{fig_m25}. Results concerning the second grid (i.e. grid (2) in Table \ref{tab_model}) are shown in Table \ref{tab_m20} and in Figures \ref{fig_z10m4} to \ref{fig_z0.01}.

Moreover, for the sake of completeness, the overproduction factors of all the s-only nuclei within the mass range $69\leq A \leq 209$ for all our s-process simulations are reported in Tables \ref{tab_overproduc_M} and \ref{tab_overproduc_Z}, in addition to a comparison with previous computations by \citet{the07} for their stellar models with a similar mass, metallicity and input physic. Other comparisons can be found in \citetalias{costa2006}.

\section{Discussion}

For all the stellar models of different initial mass and metallicity, the s-process efficiency increases when overshooting is inserted in the evolutionary computations compared with ``no-overshooting'' models, as found in \citetalias{pumo2006} and \citetalias{costa2006} for simulations referring to a 25 $M_\odot$ star model with an initial metallicity of $Z$=$0.02$. Indeed, for all our sets of models with a fixed initial mass and metallicity, a not negligible enhancement of the s-process efficiency occurs when passing from the ``no-overshooting'' model ($f=10^{-5}$) of a set to any model of the same set including overshooting, with enhancements for the main s-process indicators $F_0$ and $n_c$ until a factor $\sim 6$ and $\sim 3$, respectively.

Moreover an essentially monotonic link between the $f$ value and the s-process efficiency is evident when overshooting is inserted in the evolutionary computations, as witnessed by the fact that both main s-process indicators $F_0$ and $n_c$ (and all the s-process efficiency indicators more in general) gradually grow when passing from $f$ = 0.01 to 0.035 for any set of models with a fixed initial mass and metallicity. The only exceptions to this last behavior occur for the two models with $f=0.01$ having Z=10$^{-4}$ and Z=0.01, which give birth to a more efficient s-process than the one obtained for the corresponding models with $f=0.02$ (see set (a) and (c) of Table \ref{tab_m20}, and Fig.s \ref{fig_z10m4} and \ref{fig_z0.01}). The reason of this not strictly monotonic increase of the s-process efficiency with the $f$ value may be due to the particularly long duration of the core He-burning s-process for the two models with $f=0.01$, which leads to the enhancement of the $F_0$ and $n_c$ values and, consequently, of the s-process efficiency because of a longer neutron exposure \citepalias[see also][for details]{costa2006}. Indeed, the duration of the core He-burning s-process for the model with $f=0.01$ having Z=10$^{-4}$ is almost equal to the one of the ``no-overshooting'' model with same initial metallicity (see set (a) of Table \ref{tab_m20}), while the duration of the core He-burning s-process for model with $f=0.01$ having Z=0.01 is the longest with respect to all the other 20 $M_\odot$ stellar models of different initial metallicity with overshooting (see Table \ref{tab_m20} and set (b) of Table \ref{tab_z0.02}).

Also evident is a clear trend with both initial mass and Z (excluding two exceptions discussed below) when setting the $f$ value, according to which the s-process efficiency increases when we progressively increase both the mass of the models from 15 M$_\odot$ to 25 M$_\odot$ (see Table \ref{tab_z0.02}) and the Z value from 10$^{-4}$ to 0.02 (see Table \ref{tab_m20} and set (b) of Table \ref{tab_z0.02}), confirming the results found in other works referring to evolutionary computations without extra mixing processes owing to convective overshooting \citep[see e.g.][]{prantzos90,rayethashimoto2000,the2000,the07}. As already mentioned in Section \ref{sec_intro}, the increase in s-process efficiency with the initial mass is connected to the fact that the reaction \reac{22}{Ne}{\alpha}{n}{25}{Mg} and, consequently, the production of s-nuclei is more and more efficient when the initial stellar mass is increased; instead the enhancement of the s-process efficiency with Z is linked to the fact that the {\it source/seed} ratio is constant with Z in our models, so the efficiency increases when the Z value is increased. The only exceptions to this last trend occurs for the two models with Z=0.01 having $f$=10$^{-5}$ and $f$=0.01 (set (b) of Table \ref{tab_z0.02}), which give rise to a more efficient s-process than the one obtained for the corresponding models with Z=0.02 (set (c) of Table \ref{tab_m20}). Once again the reason for this behavior may be connected to the same previously explained effects that are related to the longer duration of the core He-burning s-process for the two models with Z=0.01 (compared to the one of the corresponding models with Z=0.02), which leads to the enhancement of the s-process efficiency.

In addition we find that the use of the NACRE rate for the \reac{12}{C}{\alpha}{\gamma}{16}{O} reaction gives rise to a lower s-process efficiency (see Figure \ref{fig_m25}). This behavior seems to be connected both to the smaller lifetime of the He-burning phase in our new $25$ M$_\odot$ models --- compared to those from \citetalias{pumo2006} and \citetalias{costa2006} --- which has a direct impact on the neutron exposure of the s-process seed (mainly \chem{56}{Fe}), and to a higher availability of $\alpha$ particles during the late He-burning phase because less $\alpha$ particles are consumed by the \reac{12}{C}{\alpha}{\gamma}{16}{O} reaction because of the lower rate, as suggested by \citet{the2000}. Moreover, although this reaction has only a relatively small influence on the efficiency of the s-process during the He-burning phase \citep[see also][]{the07}, its impact seems to be more important and, consequently, not negligible when the overshooting is inserted in stellar models. Indeed, the diminution (a factor of $\sim1.5$) in the average overproduction factor F$_0$ for the model with overshooting is higher than that (a factor of $\sim1.1$, i.e. in practice unchanged) for the model without overshooting, and a similar behavior is found for all remaining s-process efficiency indicators.

Furthermore, comparing Fig. \ref{fig_m15} to Fig. \ref{fig_m25} (see also Table \ref{tab_z0.02}), it appears that the s-nuclei production in the $Z=0.02$ models with $f\sim0.02$-$0.035$ having a given initial mass $M_{ini}$ seems to have a behavior similar to that of the $Z=0.02$ models without convective overshoot and with an initial mass $\sim M_{ini}+5$\msun. This behavior could give some indication on how the convective overshooting affects the yield of a generation of stars. Indeed, the integrated yields of stellar models with convective overshooting would be, at a first approximation, simply calculated from to the integrated yield of stellar models without overshooting, but shifted appropriately in mass. For example, to give a clue about the impact on the integrated yield of stellar models with $f\sim0.02$-$0.035$, the number of 15\msun\, stars (evaluated according a given initial mass function) should be multiplied by the yields (evaluated from models without convective overshooting) of the 20\msun\, stars and a similar shift could be used for other masses.

\section{Summary and further comments}
\label{Summary}
Many studies have been devoted to the s-process weak component so far, but a detailed scrutiny of the impact of the stellar evolution modelling uncertainties on this component is still missing. In the light of our preparatory studies on this topic (\citetalias{pumo2006} and \citetalias{costa2006}), we performed a comprehensive and quantitative study on the role of convective overshooting, considering stellar models with different initial mass and metallicity ($15 \leq M_{ZAMS}/M_{\odot} \leq 25$; $10^{-4} \leq Z \leq 0.02$).

To perform this analysis, we used the same procedure as described in \citetalias{pumo2006} and \citetalias{costa2006}. According to this procedure, stellar models have been evolved until He exhaustion in the core, using a diffusive approach to describe the convective overshooting where a so-called overshooting parameter $f$ determines the extension of the convectively mixed core and, consequently, the overall efficiency of convective overshooting. Then models have been used to ``post-process'' the stellar nuclidic composition with our s-nucleosynthesis code.

The results show that models with overshooting give a higher s-process efficiency compared with ``no-overshooting'' models for a given initial mass and metallicity. Less important but not negligible variations are clearly visible when changing the overshooting parameter $f$ value in the $0.01-0.035$ range, and the link between the $f$ value and the s-process indicators values is essentially monotonic in all our models of different initial mass and metallicity. Also evident is the higher s-process efficiency when for a given f value we progressively increase both the mass of the models from 15 M$_\odot$ to 25 M$_\odot$ and the Z value from 10$^{-4}$ to 0.02.

These results clearly show the level of uncertainty (up to a factor $\sim 6$) in the modeling of the weak s-process component due to the current lack of a self-consistent theory describing mixing processes inside the stars. Thus, prior to giving a final conclusion on the possible contribution of post-He burning phases to the s-process yields from a quantitative point of view, some additional investigation taking into account stellar evolution uncertainties in addition to the nuclear physics ones should be performed. Moreover, this additional investigation may shed light on different open questions linked, for example, to the effective existence of the LEPP process and to the model for the p-process taking place in the type II supernovae O-Ne layers, because the relevant s-nuclei are p-process seeds \citep[see e.g.][and references therein]{Arnould03}.

\begin{acknowledgements}
M.L.P. acknowledges the support by the Padua municipality {\it ``Padova Citt\`a delle Stelle''} prize and by the Bonino-Pulejo Foundation. We also warmly thank the referee G. Meynet for his valuable suggestions to improve our manuscript.
\end{acknowledgements}

\bibliographystyle{aa}

\Online
\begin{appendix}
\section{Online material}
\label{app1}

\addtocounter{table}{3} 
\longtabL{4}{
\begin{landscape}
\begin{longtable}{l|ccccc|ccccc|ccc}
\caption{\label{tab_overproduc_M} Overproduction factor of the 49 s-nuclei\tablefootmark{*} within the mass range $69\leq A \leq 209$, for all the models with an initial metallicity $Z$=0.02.}\\

\hline\hline
\multicolumn{1}{l}{} & \multicolumn{5}{c|}{15\msun} & \multicolumn{5}{c|}{20\msun} & \multicolumn{3}{c}{25\msun} \\
\hline
\multicolumn{1}{l}{} & T07 & 0.00001 & 0.01 & 0.02 & 0.035 & T07 & 0.00001 & 0.01 & 0.02 & 0.035 & T07 & 0.00001 & 0.01 \\
\hline
\endfirsthead
\caption{continued.}\\
\hline\hline
\multicolumn{1}{l}{} & \multicolumn{5}{c|}{15\msun} & \multicolumn{5}{c|}{20\msun} & \multicolumn{3}{c}{25\msun} \\
\hline
\multicolumn{1}{l}{} & T07 & 0.00001 & 0.01 & 0.02 & 0.035 & T07 & 0.00001 & 0.01 & 0.02 & 0.035 & T07 & 0.00001 & 0.01 \\
\hline
\endhead
\hline
\endfoot

$\!\!\!$\chem{69}{Ga} $\!\!\!$&....      $\!\!$& 4.97\,(+0)$\!\!\!\!$& 1.03\,(+1)$\!\!\!\!$& 2.33\,(+1)$\!\!\!\!$& 5.09\,(+1)$\!\!$&....      $\!\!$& 4.00\,(+1)$\!\!\!\!$& 4.50\,(+1)$\!\!\!\!$& 7.75\,(+1)$\!\!\!\!$& 1.27\,(+2)$\!\!$ &....      $\!\!$& 7.97\,(+1)$\!\!\!\!$& 1.23\,(+2)$\!\!$\\
$\!\!\!$\chem{71}{Ga} $\!\!\!$&....      $\!\!$& 5.77\,(+0)$\!\!\!\!$& 1.09\,(+1)$\!\!\!\!$& 2.42\,(+1)$\!\!\!\!$& 5.57\,(+1)$\!\!$&....      $\!\!$& 4.29\,(+1)$\!\!\!\!$& 4.88\,(+1)$\!\!\!\!$& 8.84\,(+1)$\!\!\!\!$& 1.54\,(+2)$\!\!$ &....      $\!\!$& 9.12\,(+1)$\!\!\!\!$& 1.48\,(+2)$\!\!$\\   
$\!\!\!$\chem{70}{Ge} $\!\!\!$&9.83\,(+0)$\!\!$& 6.02\,(+0)$\!\!\!\!$& 1.19\,(+1)$\!\!\!\!$& 2.66\,(+1)$\!\!\!\!$& 5.96\,(+1)$\!\!$&4.26\,(+1)$\!\!$& 4.64\,(+1)$\!\!\!\!$& 5.24\,(+1)$\!\!\!\!$& 9.26\,(+1)$\!\!\!\!$& 1.56\,(+2)$\!\!$ &1.07\,(+2)$\!\!$& 9.51\,(+1)$\!\!\!\!$& 1.50\,(+2)$\!\!$\\
$\!\!\!$\chem{72}{Ge} $\!\!\!$&5.96\,(+0)$\!\!$& 2.50\,(+0)$\!\!\!\!$& 4.54\,(+0)$\!\!\!\!$& 9.83\,(+0)$\!\!\!\!$& 2.28\,(+1)$\!\!$&2.60\,(+1)$\!\!$& 1.74\,(+1)$\!\!\!\!$& 1.98\,(+1)$\!\!\!\!$& 3.67\,(+1)$\!\!\!\!$& 6.51\,(+1)$\!\!$ &7.22\,(+1)$\!\!$& 3.78\,(+1)$\!\!\!\!$& 6.25\,(+1)$\!\!$\\
$\!\!\!$\chem{76}{Se} $\!\!\!$&6.01\,(+0)$\!\!$& 6.11\,(+0)$\!\!\!\!$& 1.05\,(+1)$\!\!\!\!$& 2.05\,(+1)$\!\!\!\!$& 4.71\,(+1)$\!\!$&2.46\,(+1)$\!\!$& 3.58\,(+1)$\!\!\!\!$& 4.08\,(+1)$\!\!\!\!$& 7.90\,(+1)$\!\!\!\!$& 1.52\,(+2)$\!\!$ &7.47\,(+1)$\!\!$& 8.20\,(+1)$\!\!\!\!$& 1.45\,(+2)$\!\!$\\
$\!\!\!$\chem{80}{Kr} $\!\!\!$&1.52\,(+1)$\!\!$& 1.34\,(+1)$\!\!\!\!$& 2.29\,(+1)$\!\!\!\!$& 4.18\,(+1)$\!\!\!\!$& 9.00\,(+1)$\!\!$&5.57\,(+1)$\!\!$& 6.91\,(+1)$\!\!\!\!$& 7.85\,(+1)$\!\!\!\!$& 1.52\,(+2)$\!\!\!\!$& 3.03\,(+2)$\!\!$ &1.74\,(+2)$\!\!$& 1.57\,(+2)$\!\!\!\!$& 2.89\,(+2)$\!\!$\\
$\!\!\!$\chem{82}{Kr} $\!\!\!$&7.83\,(+0)$\!\!$& 7.16\,(+0)$\!\!\!\!$& 1.12\,(+1)$\!\!\!\!$& 1.97\,(+1)$\!\!\!\!$& 4.04\,(+1)$\!\!$&2.36\,(+1)$\!\!$& 3.15\,(+1)$\!\!\!\!$& 3.55\,(+1)$\!\!\!\!$& 6.71\,(+1)$\!\!\!\!$& 1.35\,(+2)$\!\!$ &7.34\,(+1)$\!\!$& 6.94\,(+1)$\!\!\!\!$& 1.29\,(+2)$\!\!$\\
$\!\!\!$\chem{86}{Kr} $\!\!\!$&9.40\,(-1)$\!\!$& 8.37\,(-1)$\!\!\!\!$& 7.77\,(-1)$\!\!\!\!$& 7.24\,(-1)$\!\!\!\!$& 6.82\,(-1)$\!\!$&1.34\,(+0)$\!\!$& 7.07\,(-1)$\!\!\!\!$& 6.90\,(-1)$\!\!\!\!$& 6.88\,(-1)$\!\!\!\!$& 7.94\,(-1)$\!\!$ &2.57\,(+0)$\!\!$& 7.43\,(-1)$\!\!\!\!$& 9.71\,(-1)$\!\!$\\
$\!\!\!$\chem{87}{Rb} $\!\!\!$&7.20\,(-1)$\!\!$& 7.87\,(-1)$\!\!\!\!$& 7.41\,(-1)$\!\!\!\!$& 6.88\,(-1)$\!\!\!\!$& 6.52\,(-1)$\!\!$&8.40\,(-1)$\!\!$& 6.71\,(-1)$\!\!\!\!$& 6.42\,(-1)$\!\!\!\!$& 6.22\,(-1)$\!\!\!\!$& 6.43\,(-1)$\!\!$ &1.26\,(+0)$\!\!$& 6.36\,(-1)$\!\!\!\!$& 6.85\,(-1)$\!\!$\\
$\!\!\!$\chem{86}{Sr} $\!\!\!$&1.20\,(+1)$\!\!$& 1.36\,(+1)$\!\!\!\!$& 1.88\,(+1)$\!\!\!\!$& 2.94\,(+1)$\!\!\!\!$& 5.34\,(+1)$\!\!$&2.26\,(+1)$\!\!$& 4.33\,(+1)$\!\!\!\!$& 4.79\,(+1)$\!\!\!\!$& 8.24\,(+1)$\!\!\!\!$& 1.60\,(+2)$\!\!$ &5.70\,(+1)$\!\!$& 8.47\,(+1)$\!\!\!\!$& 1.52\,(+2)$\!\!$\\
$\!\!\!$\chem{87}{Sr} $\!\!\!$&1.11\,(+1)$\!\!$& 1.23\,(+1)$\!\!\!\!$& 1.71\,(+1)$\!\!\!\!$& 2.57\,(+1)$\!\!\!\!$& 4.49\,(+1)$\!\!$&2.03\,(+1)$\!\!$& 3.68\,(+1)$\!\!\!\!$& 4.05\,(+1)$\!\!\!\!$& 6.76\,(+1)$\!\!\!\!$& 1.28\,(+2)$\!\!$ &4.73\,(+1)$\!\!$& 6.92\,(+1)$\!\!\!\!$& 1.22\,(+2)$\!\!$\\
$\!\!\!$\chem{88}{Sr} $\!\!\!$&3.70\,(+0)$\!\!$& 3.97\,(+0)$\!\!\!\!$& 6.30\,(+0)$\!\!\!\!$& 9.62\,(+0)$\!\!\!\!$& 1.55\,(+1)$\!\!$&7.53\,(+0)$\!\!$& 1.31\,(+1)$\!\!\!\!$& 1.42\,(+1)$\!\!\!\!$& 2.15\,(+1)$\!\!\!\!$& 3.62\,(+1)$\!\!$ &1.41\,(+1)$\!\!$& 2.19\,(+1)$\!\!\!\!$& 3.45\,(+1)$\!\!$\\
$\!\!\!$\chem{89}{Y}  $\!\!\!$&....      $\!\!$& 1.00\,(+0)$\!\!\!\!$& 1.00\,(+0)$\!\!\!\!$& 1.00\,(+0)$\!\!\!\!$& 1.01\,(+0)$\!\!$&....      $\!\!$& 1.02\,(+1)$\!\!\!\!$& 1.11\,(+1)$\!\!\!\!$& 1.63\,(+1)$\!\!\!\!$& 2.60\,(+1)$\!\!$ &....      $\!\!$& 1.66\,(+1)$\!\!\!\!$& 2.48\,(+1)$\!\!$\\
$\!\!\!$\chem{90}{Zr} $\!\!\!$&....      $\!\!$& 1.31\,(+0)$\!\!\!\!$& 1.98\,(+0)$\!\!\!\!$& 3.19\,(+0)$\!\!\!\!$& 5.18\,(+0)$\!\!$&....      $\!\!$& 4.41\,(+0)$\!\!\!\!$& 4.76\,(+0)$\!\!\!\!$& 6.96\,(+0)$\!\!\!\!$& 1.06\,(+1)$\!\!$ &....      $\!\!$& 7.02\,(+0)$\!\!\!\!$& 1.01\,(+1)$\!\!$\\
$\!\!\!$\chem{91}{Zr} $\!\!\!$&....      $\!\!$& 1.61\,(+0)$\!\!\!\!$& 2.31\,(+0)$\!\!\!\!$& 3.69\,(+0)$\!\!\!\!$& 6.10\,(+0)$\!\!$&....      $\!\!$& 5.17\,(+0)$\!\!\!\!$& 5.61\,(+0)$\!\!\!\!$& 8.29\,(+0)$\!\!\!\!$& 1.28\,(+1)$\!\!$ &....      $\!\!$& 8.44\,(+0)$\!\!\!\!$& 1.22\,(+1)$\!\!$\\
$\!\!\!$\chem{92}{Zr} $\!\!\!$&....      $\!\!$& 1.63\,(+0)$\!\!\!\!$& 2.10\,(+0)$\!\!\!\!$& 3.12\,(+0)$\!\!\!\!$& 5.06\,(+0)$\!\!$&....      $\!\!$& 4.30\,(+0)$\!\!\!\!$& 4.66\,(+0)$\!\!\!\!$& 6.86\,(+0)$\!\!\!\!$& 1.04\,(+1)$\!\!$ &....      $\!\!$& 6.98\,(+0)$\!\!\!\!$& 1.00\,(+1)$\!\!$\\
$\!\!\!$\chem{94}{Zr} $\!\!\!$&....      $\!\!$& 1.49\,(+0)$\!\!\!\!$& 1.86\,(+0)$\!\!\!\!$& 2.46\,(+0)$\!\!\!\!$& 3.65\,(+0)$\!\!$&....      $\!\!$& 3.19\,(+0)$\!\!\!\!$& 3.40\,(+0)$\!\!\!\!$& 4.83\,(+0)$\!\!\!\!$& 7.22\,(+0)$\!\!$ &....      $\!\!$& 4.91\,(+0)$\!\!\!\!$& 6.95\,(+0)$\!\!$\\
$\!\!\!$\chem{96}{Mo} $\!\!\!$&....      $\!\!$& 1.49\,(+0)$\!\!\!\!$& 1.81\,(+0)$\!\!\!\!$& 2.34\,(+0)$\!\!\!\!$& 3.35\,(+0)$\!\!$&....      $\!\!$& 2.96\,(+0)$\!\!\!\!$& 3.13\,(+0)$\!\!\!\!$& 4.37\,(+0)$\!\!\!\!$& 6.50\,(+0)$\!\!$ &....      $\!\!$& 4.44\,(+0)$\!\!\!\!$& 6.25\,(+0)$\!\!$\\
$\!\!\!$\chem{100}{Ru}$\!\!\!$&....      $\!\!$& 1.55\,(+0)$\!\!\!\!$& 1.67\,(+0)$\!\!\!\!$& 2.06\,(+0)$\!\!\!\!$& 2.77\,(+0)$\!\!$&....      $\!\!$& 2.50\,(+0)$\!\!\!\!$& 2.62\,(+0)$\!\!\!\!$& 3.50\,(+0)$\!\!\!\!$& 5.05\,(+0)$\!\!$ &....      $\!\!$& 3.54\,(+0)$\!\!\!\!$& 4.86\,(+0)$\!\!$\\
$\!\!\!$\chem{104}{Pd}$\!\!\!$&....      $\!\!$& 1.78\,(+0)$\!\!\!\!$& 1.73\,(+0)$\!\!\!\!$& 2.05\,(+0)$\!\!\!\!$& 2.66\,(+0)$\!\!$&....      $\!\!$& 2.46\,(+0)$\!\!\!\!$& 2.54\,(+0)$\!\!\!\!$& 3.30\,(+0)$\!\!\!\!$& 4.67\,(+0)$\!\!$ &....      $\!\!$& 3.35\,(+0)$\!\!\!\!$& 4.50\,(+0)$\!\!$\\
$\!\!\!$\chem{110}{Cd}$\!\!\!$&....      $\!\!$& 1.78\,(+0)$\!\!\!\!$& 1.56\,(+0)$\!\!\!\!$& 1.67\,(+0)$\!\!\!\!$& 2.09\,(+0)$\!\!$&....      $\!\!$& 1.95\,(+0)$\!\!\!\!$& 2.01\,(+0)$\!\!\!\!$& 2.51\,(+0)$\!\!\!\!$& 3.43\,(+0)$\!\!$ &....      $\!\!$& 2.55\,(+0)$\!\!\!\!$& 3.32\,(+0)$\!\!$\\
$\!\!\!$\chem{116}{Sn}$\!\!\!$&....      $\!\!$& 1.98\,(+0)$\!\!\!\!$& 1.72\,(+0)$\!\!\!\!$& 1.55\,(+0)$\!\!\!\!$& 1.70\,(+0)$\!\!$&....      $\!\!$& 1.62\,(+0)$\!\!\!\!$& 1.66\,(+0)$\!\!\!\!$& 1.94\,(+0)$\!\!\!\!$& 2.48\,(+0)$\!\!$ &....      $\!\!$& 1.96\,(+0)$\!\!\!\!$& 2.42\,(+0)$\!\!$\\
$\!\!\!$\chem{118}{Sn}$\!\!\!$&....      $\!\!$& 1.88\,(+0)$\!\!\!\!$& 1.66\,(+0)$\!\!\!\!$& 1.45\,(+0)$\!\!\!\!$& 1.42\,(+0)$\!\!$&....      $\!\!$& 1.40\,(+0)$\!\!\!\!$& 1.41\,(+0)$\!\!\!\!$& 1.53\,(+0)$\!\!\!\!$& 1.86\,(+0)$\!\!$ &....      $\!\!$& 1.55\,(+0)$\!\!\!\!$& 1.83\,(+0)$\!\!$\\
$\!\!\!$\chem{122}{Te}$\!\!\!$&....      $\!\!$& 2.47\,(+0)$\!\!\!\!$& 2.76\,(+0)$\!\!\!\!$& 2.62\,(+0)$\!\!\!\!$& 2.39\,(+0)$\!\!$&....      $\!\!$& 2.45\,(+0)$\!\!\!\!$& 2.42\,(+0)$\!\!\!\!$& 2.32\,(+0)$\!\!\!\!$& 2.46\,(+0)$\!\!$ &....      $\!\!$& 2.31\,(+0)$\!\!\!\!$& 2.44\,(+0)$\!\!$\\
$\!\!\!$\chem{124}{Te}$\!\!\!$&....      $\!\!$& 2.38\,(+0)$\!\!\!\!$& 2.79\,(+0)$\!\!\!\!$& 2.75\,(+0)$\!\!\!\!$& 2.52\,(+0)$\!\!$&....      $\!\!$& 2.58\,(+0)$\!\!\!\!$& 2.56\,(+0)$\!\!\!\!$& 2.42\,(+0)$\!\!\!\!$& 2.49\,(+0)$\!\!$ &....      $\!\!$& 2.40\,(+0)$\!\!\!\!$& 2.47\,(+0)$\!\!$\\
$\!\!\!$\chem{128}{Xe}$\!\!\!$&....      $\!\!$& 2.04\,(+0)$\!\!\!\!$& 2.57\,(+0)$\!\!\!\!$& 2.79\,(+0)$\!\!\!\!$& 2.65\,(+0)$\!\!$&....      $\!\!$& 2.75\,(+0)$\!\!\!\!$& 2.70\,(+0)$\!\!\!\!$& 2.52\,(+0)$\!\!\!\!$& 2.42\,(+0)$\!\!$ &....      $\!\!$& 2.51\,(+0)$\!\!\!\!$& 2.42\,(+0)$\!\!$\\
$\!\!\!$\chem{130}{Xe}$\!\!\!$&....      $\!\!$& 2.14\,(+0)$\!\!\!\!$& 2.52\,(+0)$\!\!\!\!$& 2.83\,(+0)$\!\!\!\!$& 2.73\,(+0)$\!\!$&....      $\!\!$& 2.86\,(+0)$\!\!\!\!$& 2.78\,(+0)$\!\!\!\!$& 2.60\,(+0)$\!\!\!\!$& 2.46\,(+0)$\!\!$ &....      $\!\!$& 2.59\,(+0)$\!\!\!\!$& 2.45\,(+0)$\!\!$\\
$\!\!\!$\chem{134}{Ba}$\!\!\!$&....      $\!\!$& 4.85\,(+0)$\!\!\!\!$& 4.40\,(+0)$\!\!\!\!$& 4.95\,(+0)$\!\!\!\!$& 5.30\,(+0)$\!\!$&....      $\!\!$& 5.27\,(+0)$\!\!\!\!$& 5.29\,(+0)$\!\!\!\!$& 5.21\,(+0)$\!\!\!\!$& 4.90\,(+0)$\!\!$ &....      $\!\!$& 5.17\,(+0)$\!\!\!\!$& 4.88\,(+0)$\!\!$\\
$\!\!\!$\chem{136}{Ba}$\!\!\!$&....      $\!\!$& 5.32\,(+0)$\!\!\!\!$& 4.16\,(+0)$\!\!\!\!$& 4.07\,(+0)$\!\!\!\!$& 4.43\,(+0)$\!\!$&....      $\!\!$& 4.29\,(+0)$\!\!\!\!$& 4.38\,(+0)$\!\!\!\!$& 4.51\,(+0)$\!\!\!\!$& 4.40\,(+0)$\!\!$ &....      $\!\!$& 4.49\,(+0)$\!\!\!\!$& 4.40\,(+0)$\!\!$\\
$\!\!\!$\chem{137}{Ba}$\!\!\!$&....      $\!\!$& 3.25\,(+0)$\!\!\!\!$& 2.62\,(+0)$\!\!\!\!$& 2.32\,(+0)$\!\!\!\!$& 2.43\,(+0)$\!\!$&....      $\!\!$& 2.35\,(+0)$\!\!\!\!$& 2.40\,(+0)$\!\!\!\!$& 2.51\,(+0)$\!\!\!\!$& 2.51\,(+0)$\!\!$ &....      $\!\!$& 2.50\,(+0)$\!\!\!\!$& 2.51\,(+0)$\!\!$\\
$\!\!\!$\chem{138}{Ba}$\!\!\!$&....      $\!\!$& 2.47\,(+0)$\!\!\!\!$& 3.12\,(+0)$\!\!\!\!$& 3.57\,(+0)$\!\!\!\!$& 3.98\,(+0)$\!\!$&....      $\!\!$& 3.81\,(+0)$\!\!\!\!$& 3.91\,(+0)$\!\!\!\!$& 4.23\,(+0)$\!\!\!\!$& 4.57\,(+0)$\!\!$ &....      $\!\!$& 4.24\,(+0)$\!\!\!\!$& 4.54\,(+0)$\!\!$\\
$\!\!\!$\chem{139}{La}$\!\!\!$&....      $\!\!$& 1.37\,(+0)$\!\!\!\!$& 1.88\,(+0)$\!\!\!\!$& 2.32\,(+0)$\!\!\!\!$& 2.69\,(+0)$\!\!$&....      $\!\!$& 2.56\,(+0)$\!\!\!\!$& 2.63\,(+0)$\!\!\!\!$& 2.89\,(+0)$\!\!\!\!$& 3.17\,(+0)$\!\!$ &....      $\!\!$& 2.91\,(+0)$\!\!\!\!$& 3.15\,(+0)$\!\!$\\
$\!\!\!$\chem{140}{Ce}$\!\!\!$&....      $\!\!$& 1.34\,(+0)$\!\!\!\!$& 1.66\,(+0)$\!\!\!\!$& 2.11\,(+0)$\!\!\!\!$& 2.63\,(+0)$\!\!$&....      $\!\!$& 2.45\,(+0)$\!\!\!\!$& 2.54\,(+0)$\!\!\!\!$& 2.96\,(+0)$\!\!\!\!$& 3.41\,(+0)$\!\!$ &....      $\!\!$& 2.98\,(+0)$\!\!\!\!$& 3.37\,(+0)$\!\!$\\
$\!\!\!$\chem{142}{Nd}$\!\!\!$&....      $\!\!$& 1.59\,(+0)$\!\!\!\!$& 1.82\,(+0)$\!\!\!\!$& 2.23\,(+0)$\!\!\!\!$& 2.83\,(+0)$\!\!$&....      $\!\!$& 2.61\,(+0)$\!\!\!\!$& 2.72\,(+0)$\!\!\!\!$& 3.24\,(+0)$\!\!\!\!$& 3.85\,(+0)$\!\!$ &....      $\!\!$& 3.27\,(+0)$\!\!\!\!$& 3.80\,(+0)$\!\!$\\
$\!\!\!$\chem{148}{Sm}$\!\!\!$&....      $\!\!$& 1.94\,(+0)$\!\!\!\!$& 1.94\,(+0)$\!\!\!\!$& 2.13\,(+0)$\!\!\!\!$& 2.55\,(+0)$\!\!$&....      $\!\!$& 2.39\,(+0)$\!\!\!\!$& 2.46\,(+0)$\!\!\!\!$& 2.92\,(+0)$\!\!\!\!$& 3.52\,(+0)$\!\!$ &....      $\!\!$& 2.95\,(+0)$\!\!\!\!$& 3.46\,(+0)$\!\!$\\
$\!\!\!$\chem{150}{Sm}$\!\!\!$&....      $\!\!$& 1.70\,(+0)$\!\!\!\!$& 1.68\,(+0)$\!\!\!\!$& 1.82\,(+0)$\!\!\!\!$& 2.17\,(+0)$\!\!$&....      $\!\!$& 2.05\,(+0)$\!\!\!\!$& 2.10\,(+0)$\!\!\!\!$& 2.49\,(+0)$\!\!\!\!$& 3.00\,(+0)$\!\!$ &....      $\!\!$& 2.51\,(+0)$\!\!\!\!$& 2.95\,(+0)$\!\!$\\
$\!\!\!$\chem{154}{Gd}$\!\!\!$&....      $\!\!$& 1.87\,(+0)$\!\!\!\!$& 1.82\,(+0)$\!\!\!\!$& 1.96\,(+0)$\!\!\!\!$& 2.31\,(+0)$\!\!$&....      $\!\!$& 2.20\,(+0)$\!\!\!\!$& 2.24\,(+0)$\!\!\!\!$& 2.65\,(+0)$\!\!\!\!$& 3.20\,(+0)$\!\!$ &....      $\!\!$& 2.69\,(+0)$\!\!\!\!$& 3.15\,(+0)$\!\!$\\
$\!\!\!$\chem{160}{Dy}$\!\!\!$&....      $\!\!$& 1.78\,(+0)$\!\!\!\!$& 1.69\,(+0)$\!\!\!\!$& 1.77\,(+0)$\!\!\!\!$& 2.05\,(+0)$\!\!$&....      $\!\!$& 2.03\,(+0)$\!\!\!\!$& 2.02\,(+0)$\!\!\!\!$& 2.38\,(+0)$\!\!\!\!$& 2.86\,(+0)$\!\!$ &....      $\!\!$& 2.44\,(+0)$\!\!\!\!$& 2.81\,(+0)$\!\!$\\
$\!\!\!$\chem{170}{Yb}$\!\!\!$&....      $\!\!$& 2.97\,(+0)$\!\!\!\!$& 2.65\,(+0)$\!\!\!\!$& 2.69\,(+0)$\!\!\!\!$& 3.00\,(+0)$\!\!$&....      $\!\!$& 2.97\,(+0)$\!\!\!\!$& 2.95\,(+0)$\!\!\!\!$& 3.41\,(+0)$\!\!\!\!$& 4.15\,(+0)$\!\!$ &....      $\!\!$& 3.49\,(+0)$\!\!\!\!$& 4.11\,(+0)$\!\!$\\
$\!\!\!$\chem{176}{Hf}$\!\!\!$&....      $\!\!$& 4.02\,(+0)$\!\!\!\!$& 3.33\,(+0)$\!\!\!\!$& 3.11\,(+0)$\!\!\!\!$& 3.23\,(+0)$\!\!$&....      $\!\!$& 3.19\,(+0)$\!\!\!\!$& 3.18\,(+0)$\!\!\!\!$& 3.50\,(+0)$\!\!\!\!$& 4.01\,(+0)$\!\!$ &....      $\!\!$& 3.47\,(+0)$\!\!\!\!$& 3.90\,(+0)$\!\!$\\
$\!\!\!$\chem{186}{Os}$\!\!\!$&....      $\!\!$& 7.48\,(+0)$\!\!\!\!$& 3.81\,(+0)$\!\!\!\!$& 3.03\,(+0)$\!\!\!\!$& 2.94\,(+0)$\!\!$&....      $\!\!$& 2.90\,(+0)$\!\!\!\!$& 2.92\,(+0)$\!\!\!\!$& 3.05\,(+0)$\!\!\!\!$& 3.47\,(+0)$\!\!$ &....      $\!\!$& 3.07\,(+0)$\!\!\!\!$& 3.46\,(+0)$\!\!$\\
$\!\!\!$\chem{187}{Os}$\!\!\!$&....      $\!\!$& 3.76\,(+0)$\!\!\!\!$& 1.92\,(+0)$\!\!\!\!$& 1.52\,(+0)$\!\!\!\!$& 1.47\,(+0)$\!\!$&....      $\!\!$& 1.45\,(+0)$\!\!\!\!$& 1.47\,(+0)$\!\!\!\!$& 1.53\,(+0)$\!\!\!\!$& 1.74\,(+0)$\!\!$ &....      $\!\!$& 1.55\,(+0)$\!\!\!\!$& 1.74\,(+0)$\!\!$\\
$\!\!\!$\chem{192}{Pt}$\!\!\!$&....      $\!\!$& 8.60\,(+0)$\!\!\!\!$& 4.52\,(+0)$\!\!\!\!$& 3.02\,(+0)$\!\!\!\!$& 2.67\,(+0)$\!\!$&....      $\!\!$& 2.87\,(+0)$\!\!\!\!$& 2.72\,(+0)$\!\!\!\!$& 2.73\,(+0)$\!\!\!\!$& 2.91\,(+0)$\!\!$ &....      $\!\!$& 2.77\,(+0)$\!\!\!\!$& 2.90\,(+0)$\!\!$\\
$\!\!\!$\chem{198}{Hg}$\!\!\!$&....      $\!\!$& 5.92\,(+0)$\!\!\!\!$& 6.37\,(+0)$\!\!\!\!$& 3.55\,(+0)$\!\!\!\!$& 2.56\,(+0)$\!\!$&....      $\!\!$& 2.82\,(+0)$\!\!\!\!$& 2.63\,(+0)$\!\!\!\!$& 2.42\,(+0)$\!\!\!\!$& 2.43\,(+0)$\!\!$ &....      $\!\!$& 2.41\,(+0)$\!\!\!\!$& 2.43\,(+0)$\!\!$\\
$\!\!\!$\chem{200}{Hg}$\!\!\!$&....      $\!\!$& 3.31\,(+0)$\!\!\!\!$& 4.39\,(+0)$\!\!\!\!$& 2.94\,(+0)$\!\!\!\!$& 1.86\,(+0)$\!\!$&....      $\!\!$& 2.06\,(+0)$\!\!\!\!$& 1.95\,(+0)$\!\!\!\!$& 1.64\,(+0)$\!\!\!\!$& 1.57\,(+0)$\!\!$ &....      $\!\!$& 1.60\,(+0)$\!\!\!\!$& 1.58\,(+0)$\!\!$\\
$\!\!\!$\chem{201}{Hg}$\!\!\!$&....      $\!\!$& 2.23\,(+0)$\!\!\!\!$& 2.92\,(+0)$\!\!\!\!$& 2.13\,(+0)$\!\!\!\!$& 1.33\,(+0)$\!\!$&....      $\!\!$& 1.47\,(+0)$\!\!\!\!$& 1.40\,(+0)$\!\!\!\!$& 1.15\,(+0)$\!\!\!\!$& 1.10\,(+0)$\!\!$ &....      $\!\!$& 1.13\,(+0)$\!\!\!\!$& 1.11\,(+0)$\!\!$\\
$\!\!\!$\chem{202}{Hg}$\!\!\!$&....      $\!\!$& 6.14\,(+0)$\!\!\!\!$& 5.36\,(+0)$\!\!\!\!$& 4.83\,(+0)$\!\!\!\!$& 3.11\,(+0)$\!\!$&....      $\!\!$& 3.55\,(+0)$\!\!\!\!$& 3.34\,(+0)$\!\!\!\!$& 2.45\,(+0)$\!\!\!\!$& 2.07\,(+0)$\!\!$ &....      $\!\!$& 2.35\,(+0)$\!\!\!\!$& 2.08\,(+0)$\!\!$\\
$\!\!\!$\chem{204}{Pb}$\!\!\!$&....      $\!\!$& 9.97\,(+0)$\!\!\!\!$& 7.49\,(+0)$\!\!\!\!$& 6.83\,(+0)$\!\!\!\!$& 5.26\,(+0)$\!\!$&....      $\!\!$& 5.79\,(+0)$\!\!\!\!$& 5.58\,(+0)$\!\!\!\!$& 4.09\,(+0)$\!\!\!\!$& 3.11\,(+0)$\!\!$ &....      $\!\!$& 3.94\,(+0)$\!\!\!\!$& 3.16\,(+0)$\!\!$\\
$\!\!\!$\chem{209}{Bi}$\!\!\!$&....      $\!\!$& 1.00\,(+0)$\!\!\!\!$& 1.02\,(+0)$\!\!\!\!$& 1.06\,(+0)$\!\!\!\!$& 1.14\,(+0)$\!\!$&....      $\!\!$& 1.11\,(+0)$\!\!\!\!$& 1.12\,(+0)$\!\!\!\!$& 1.20\,(+0)$\!\!\!\!$& 1.30\,(+0)$\!\!$ &....      $\!\!$& 1.22\,(+0)$\!\!\!\!$& 1.32\,(+0)$\!\!$\\
\hline
\end{longtable}
\tablefoot{The initial mass of the models and the value of the overshooting parameter used in the evolutionary computations are reported in the first and in the second row, respectively. The columns labeled as T07 refer to the values of overproduction factors obtained by \citet{the07} for their Z=0.02 stellar models, which are comparable with our models without overshooting. The sign ``....'' means datum not available. The notation X.XX\,($\pm$Y) has its standard meaning of X.XX $\times$ 10$^{\pm Y}$.\\ \tablefoottext{*}{These are the nuclei indicated as ``mainly'' produced by the s-process in \citet{anders_grevesse89}.}}
\end{landscape}
}


\longtabL{5}{
\begin{landscape}
\begin{longtable}{l|cccc|cccc|cccc}
\caption{\label{tab_overproduc_Z} As in Table \ref{tab_overproduc_M}, but for all the $M$=20\msun\, models with an initial metallicity $Z<$ 0.02.}\\

\hline\hline
\multicolumn{1}{l}{} & \multicolumn{4}{c|}{0.0001} & \multicolumn{4}{c|}{0.005} & \multicolumn{4}{c}{0.01} \\
\hline
\multicolumn{1}{l}{} & 0.00001 & 0.01 & 0.02 & 0.035 & 0.00001 & 0.01 & 0.02 & 0.035 & 0.00001 & 0.01 & 0.02 & 0.035\\
\hline
\endfirsthead
\caption{continued.}\\
\hline\hline
\multicolumn{1}{l}{} & \multicolumn{4}{c|}{0.0001} & \multicolumn{4}{c|}{0.005} & \multicolumn{4}{c}{0.01} \\
\hline
\multicolumn{1}{l}{} & 0.00001 & 0.01 & 0.02 & 0.035 & 0.00001 & 0.01 & 0.02 & 0.035 & 0.00001 & 0.01 & 0.02 & 0.035\\
\hline
\endhead
\hline
\endfoot

\chem{69}{Ga} &1.77\,(+0)&1.88\,(+0)&1.87\,(+0)&1.88\,(+0)&4.89\,(+0)&5.60\,(+0)&6.01\,(+0)&6.00\,(+0)&4.01\,(+1)&6.39\,(+1)&6.07\,(+1)&1.01\,(+2)\\
\chem{71}{Ga} &1.62\,(+0)&1.95\,(+0)&1.93\,(+0)&1.98\,(+0)&5.63\,(+0)&6.36\,(+0)&6.86\,(+0)&6.90\,(+0)&4.33\,(+1)&7.19\,(+1)&6.84\,(+1)&1.18\,(+2)\\
\chem{70}{Ge} &1.95\,(+0)&2.21\,(+0)&2.19\,(+0)&2.22\,(+0)&5.82\,(+0)&6.58\,(+0)&7.07\,(+0)&7.12\,(+0)&4.62\,(+1)&7.54\,(+1)&7.20\,(+1)&1.22\,(+2)\\
\chem{72}{Ge} &7.58\,(-1)&8.18\,(-1)&8.11\,(-1)&8.23\,(-1)&2.44\,(+0)&2.72\,(+0)&2.90\,(+0)&2.90\,(+0)&1.76\,(+1)&2.95\,(+1)&2.79\,(+1)&4.94\,(+1)\\
\chem{76}{Se} &2.65\,(+0)&2.76\,(+0)&2.75\,(+0)&2.76\,(+0)&5.98\,(+0)&6.70\,(+0)&7.11\,(+0)&6.98\,(+0)&3.63\,(+1)&6.26\,(+1)&5.88\,(+1)&1.11\,(+2)\\
\chem{80}{Kr} &4.84\,(+0)&5.63\,(+0)&5.58\,(+0)&5.70\,(+0)&1.33\,(+1)&1.48\,(+1)&1.57\,(+1)&1.61\,(+1)&7.00\,(+1)&1.19\,(+2)&1.12\,(+2)&2.17\,(+2)\\
\chem{82}{Kr} &3.73\,(+0)&4.11\,(+0)&4.09\,(+0)&4.15\,(+0)&6.90\,(+0)&7.61\,(+0)&7.93\,(+0)&8.02\,(+0)&3.17\,(+1)&5.30\,(+1)&5.02\,(+1)&9.60\,(+1)\\
\chem{86}{Kr} &9.82\,(-1)&9.68\,(-1)&9.67\,(-1)&9.63\,(-1)&8.41\,(-1)&8.29\,(-1)&8.17\,(-1)&8.18\,(-1)&7.42\,(-1)&7.45\,(-1)&7.05\,(-1)&7.66\,(-1)\\
\chem{87}{Rb} &9.43\,(-1)&9.28\,(-1)&9.28\,(-1)&9.33\,(-1)&8.08\,(-1)&7.98\,(-1)&7.91\,(-1)&7.84\,(-1)&6.99\,(-1)&6.52\,(-1)&6.64\,(-1)&6.61\,(-1)\\
\chem{86}{Sr} &5.91\,(+0)&6.52\,(+0)&6.48\,(+0)&6.55\,(+0)&1.34\,(+1)&1.43\,(+1)&1.50\,(+1)&1.50\,(+1)&4.35\,(+1)&6.71\,(+1)&6.35\,(+1)&1.15\,(+2)\\
\chem{87}{Sr} &3.59\,(+0)&4.59\,(+0)&4.52\,(+0)&4.64\,(+0)&1.20\,(+1)&1.28\,(+1)&1.35\,(+1)&1.37\,(+1)&3.66\,(+1)&5.55\,(+1)&5.32\,(+1)&9.25\,(+1)\\
\chem{88}{Sr} &1.15\,(+0)&1.27\,(+0)&1.26\,(+0)&1.27\,(+0)&3.88\,(+0)&4.30\,(+0)&4.57\,(+0)&4.58\,(+0)&1.30\,(+1)&1.84\,(+1)&1.77\,(+1)&2.76\,(+1)\\
\chem{89}{Y}  &1.08\,(+0)&1.13\,(+0)&1.12\,(+0)&1.13\,(+0)&2.73\,(+0)&3.04\,(+0)&3.23\,(+0)&3.24\,(+0)&1.02\,(+1)&1.42\,(+1)&1.37\,(+1)&2.05\,(+1)\\
\chem{90}{Zr} &9.73\,(-1)&9.63\,(-1)&9.64\,(-1)&9.62\,(-1)&1.28\,(+0)&1.38\,(+0)&1.43\,(+0)&1.43\,(+0)&4.36\,(+0)&6.01\,(+0)&5.83\,(+0)&8.53\,(+0)\\
\chem{91}{Zr} &1.15\,(+0)&1.21\,(+0)&1.21\,(+0)&1.22\,(+0)&1.61\,(+0)&1.73\,(+0)&1.80\,(+0)&1.80\,(+0)&5.21\,(+0)&7.25\,(+0)&6.99\,(+0)&1.02\,(+1)\\
\chem{92}{Zr} &1.12\,(+0)&1.16\,(+0)&1.16\,(+0)&1.16\,(+0)&1.62\,(+0)&1.70\,(+0)&1.74\,(+0)&1.74\,(+0)&4.31\,(+0)&5.97\,(+0)&5.74\,(+0)&8.46\,(+0)\\
\chem{94}{Zr} &8.64\,(-1)&8.76\,(-1)&8.78\,(-1)&8.83\,(-1)&1.48\,(+0)&1.54\,(+0)&1.59\,(+0)&1.59\,(+0)&3.16\,(+0)&4.23\,(+0)&4.08\,(+0)&5.88\,(+0)\\
\chem{96}{Mo} &1.89\,(+0)&1.95\,(+0)&1.94\,(+0)&1.94\,(+0)&1.51\,(+0)&1.56\,(+0)&1.60\,(+0)&1.61\,(+0)&2.95\,(+0)&3.85\,(+0)&3.74\,(+0)&5.30\,(+0)\\
\chem{100}{Ru}&1.67\,(+0)&1.70\,(+0)&1.69\,(+0)&1.70\,(+0)&1.54\,(+0)&1.55\,(+0)&1.56\,(+0)&1.56\,(+0)&2.49\,(+0)&3.11\,(+0)&3.02\,(+0)&4.17\,(+0)\\
\chem{104}{Pd}&2.47\,(+0)&2.23\,(+0)&2.25\,(+0)&2.22\,(+0)&1.81\,(+0)&1.76\,(+0)&1.70\,(+0)&1.70\,(+0)&2.49\,(+0)&2.97\,(+0)&2.90\,(+0)&3.91\,(+0)\\
\chem{110}{Cd}&2.71\,(+0)&2.81\,(+0)&2.80\,(+0)&2.80\,(+0)&1.74\,(+0)&1.69\,(+0)&1.68\,(+0)&1.66\,(+0)&1.96\,(+0)&2.30\,(+0)&2.23\,(+0)&2.90\,(+0)\\
\chem{116}{Sn}&1.57\,(+0)&1.74\,(+0)&1.73\,(+0)&1.75\,(+0)&1.91\,(+0)&1.86\,(+0)&1.87\,(+0)&1.87\,(+0)&1.61\,(+0)&1.82\,(+0)&1.79\,(+0)&2.18\,(+0)\\
\chem{118}{Sn}&1.14\,(+0)&1.21\,(+0)&1.20\,(+0)&1.21\,(+0)&1.84\,(+0)&1.81\,(+0)&1.83\,(+0)&1.83\,(+0)&1.38\,(+0)&1.48\,(+0)&1.46\,(+0)&1.68\,(+0)\\
\chem{122}{Te}&1.45\,(+0)&1.46\,(+0)&1.46\,(+0)&1.46\,(+0)&2.45\,(+0)&2.54\,(+0)&2.61\,(+0)&2.61\,(+0)&2.39\,(+0)&2.34\,(+0)&2.35\,(+0)&2.36\,(+0)\\
\chem{124}{Te}&1.61\,(+0)&1.61\,(+0)&1.61\,(+0)&1.61\,(+0)&2.36\,(+0)&2.48\,(+0)&2.56\,(+0)&2.58\,(+0)&2.52\,(+0)&2.45\,(+0)&2.47\,(+0)&2.41\,(+0)\\
\chem{128}{Xe}&2.33\,(+0)&2.04\,(+0)&2.06\,(+0)&2.03\,(+0)&2.23\,(+0)&2.28\,(+0)&2.20\,(+0)&2.20\,(+0)&2.80\,(+0)&2.57\,(+0)&2.59\,(+0)&2.46\,(+0)\\
\chem{130}{Xe}&5.08\,(+0)&4.17\,(+0)&4.24\,(+0)&4.12\,(+0)&2.33\,(+0)&2.34\,(+0)&2.20\,(+0)&2.20\,(+0)&2.96\,(+0)&2.66\,(+0)&2.69\,(+0)&2.54\,(+0)\\
\chem{134}{Ba}&7.26\,(+0)&8.16\,(+0)&8.10\,(+0)&8.19\,(+0)&4.69\,(+0)&4.50\,(+0)&4.46\,(+0)&4.45\,(+0)&5.23\,(+0)&5.24\,(+0)&5.24\,(+0)&5.04\,(+0)\\
\chem{136}{Ba}&2.99\,(+0)&3.79\,(+0)&3.73\,(+0)&3.86\,(+0)&5.05\,(+0)&4.83\,(+0)&4.88\,(+0)&4.88\,(+0)&4.23\,(+0)&4.49\,(+0)&4.48\,(+0)&4.45\,(+0)\\
\chem{137}{Ba}&1.15\,(+0)&1.37\,(+0)&1.35\,(+0)&1.39\,(+0)&3.11\,(+0)&3.03\,(+0)&3.08\,(+0)&3.08\,(+0)&2.32\,(+0)&2.48\,(+0)&2.47\,(+0)&2.52\,(+0)\\
\chem{138}{Ba}&1.10\,(+0)&1.14\,(+0)&1.14\,(+0)&1.15\,(+0)&2.44\,(+0)&2.59\,(+0)&2.70\,(+0)&2.70\,(+0)&3.75\,(+0)&4.12\,(+0)&4.08\,(+0)&4.39\,(+0)\\
\chem{139}{La}&9.30\,(-1)&9.17\,(-1)&9.18\,(-1)&9.16\,(-1)&1.39\,(+0)&1.48\,(+0)&1.53\,(+0)&1.53\,(+0)&2.54\,(+0)&2.81\,(+0)&2.78\,(+0)&3.03\,(+0)\\
\chem{140}{Ce}&1.07\,(+0)&1.09\,(+0)&1.09\,(+0)&1.09\,(+0)&1.34\,(+0)&1.39\,(+0)&1.42\,(+0)&1.42\,(+0)&2.44\,(+0)&2.81\,(+0)&2.77\,(+0)&3.18\,(+0)\\
\chem{142}{Nd}&1.34\,(+0)&1.37\,(+0)&1.37\,(+0)&1.37\,(+0)&1.58\,(+0)&1.62\,(+0)&1.65\,(+0)&1.65\,(+0)&2.60\,(+0)&3.05\,(+0)&3.00\,(+0)&3.54\,(+0)\\
\chem{148}{Sm}&2.39\,(+0)&2.43\,(+0)&2.43\,(+0)&2.43\,(+0)&1.93\,(+0)&1.92\,(+0)&1.91\,(+0)&1.91\,(+0)&2.40\,(+0)&2.74\,(+0)&2.69\,(+0)&3.21\,(+0)\\
\chem{150}{Sm}&2.21\,(+0)&2.20\,(+0)&2.20\,(+0)&2.20\,(+0)&1.73\,(+0)&1.70\,(+0)&1.68\,(+0)&1.67\,(+0)&2.08\,(+0)&2.34\,(+0)&2.29\,(+0)&2.73\,(+0)\\
\chem{154}{Gd}&2.61\,(+0)&2.47\,(+0)&2.48\,(+0)&2.46\,(+0)&2.06\,(+0)&1.97\,(+0)&1.85\,(+0)&1.82\,(+0)&2.32\,(+0)&2.51\,(+0)&2.39\,(+0)&2.93\,(+0)\\
\chem{160}{Dy}&4.61\,(+0)&3.14\,(+0)&3.20\,(+0)&3.06\,(+0)&1.98\,(+0)&1.90\,(+0)&1.81\,(+0)&1.55\,(+0)&2.28\,(+0)&2.27\,(+0)&2.00\,(+0)&2.61\,(+0)\\
\chem{170}{Yb}&1.17\,(+1)&1.17\,(+1)&1.18\,(+1)&1.18\,(+1)&3.21\,(+0)&3.10\,(+0)&2.87\,(+0)&2.98\,(+0)&3.19\,(+0)&3.28\,(+0)&3.35\,(+0)&3.83\,(+0)\\
\chem{176}{Hf}&7.83\,(+0)&9.98\,(+0)&9.78\,(+0)&1.00\,(+1)&3.79\,(+0)&3.53\,(+0)&3.50\,(+0)&3.39\,(+0)&3.20\,(+0)&3.30\,(+0)&3.15\,(+0)&3.72\,(+0)\\
\chem{186}{Os}&2.58\,(+0)&3.04\,(+0)&3.01\,(+0)&3.08\,(+0)&6.96\,(+0)&6.10\,(+0)&5.97\,(+0)&6.01\,(+0)&3.38\,(+0)&3.56\,(+0)&3.49\,(+0)&3.25\,(+0)\\
\chem{187}{Os}&1.36\,(+0)&1.53\,(+0)&1.52\,(+0)&1.55\,(+0)&3.64\,(+0)&3.20\,(+0)&3.11\,(+0)&3.14\,(+0)&1.74\,(+0)&1.83\,(+0)&1.78\,(+0)&1.63\,(+0)\\
\chem{192}{Pt}&4.65\,(+0)&3.23\,(+0)&3.28\,(+0)&3.16\,(+0)&8.90\,(+0)&7.97\,(+0)&7.35\,(+0)&7.20\,(+0)&4.16\,(+0)&4.11\,(+0)&3.84\,(+0)&2.88\,(+0)\\
\chem{198}{Hg}&1.85\,(+1)&1.88\,(+1)&1.89\,(+1)&1.87\,(+1)&6.29\,(+0)&6.86\,(+0)&6.98\,(+0)&6.96\,(+0)&3.77\,(+0)&5.66\,(+0)&5.06\,(+0)&2.47\,(+0)\\
\chem{200}{Hg}&5.93\,(+0)&8.13\,(+0)&7.98\,(+0)&8.27\,(+0)&3.30\,(+0)&3.61\,(+0)&3.81\,(+0)&3.80\,(+0)&2.25\,(+0)&3.65\,(+0)&3.20\,(+0)&1.60\,(+0)\\
\chem{201}{Hg}&2.86\,(+0)&4.36\,(+0)&4.26\,(+0)&4.49\,(+0)&2.22\,(+0)&2.39\,(+0)&2.53\,(+0)&2.56\,(+0)&1.56\,(+0)&2.48\,(+0)&2.20\,(+0)&1.12\,(+0)\\
\chem{202}{Hg}&2.28\,(+0)&3.42\,(+0)&3.31\,(+0)&3.51\,(+0)&5.60\,(+0)&5.23\,(+0)&5.33\,(+0)&5.33\,(+0)&3.39\,(+0)&3.95\,(+0)&3.64\,(+0)&2.20\,(+0)\\
\chem{204}{Pb}&1.38\,(+0)&1.65\,(+0)&1.62\,(+0)&1.67\,(+0)&9.44\,(+0)&9.03\,(+0)&9.16\,(+0)&9.12\,(+0)&5.53\,(+0)&5.10\,(+0)&5.03\,(+0)&3.47\,(+0)\\
\chem{209}{Bi}&9.87\,(-1)&9.89\,(-1)&9.89\,(-1)&9.90\,(-1)&1.03\,(+0)&1.03\,(+0)&1.03\,(+0)&1.03\,(+0)&1.13\,(+0)&1.19\,(+0)&1.17\,(+0)&1.26\,(+0)\\
\hline
\end{longtable}
\tablefoot{The initial metallicity of the models and the value of the overshooting parameter used in the evolutionary computations are reported in the first and in the second row, respectively.}
\end{landscape}
}

\end{appendix}

\end{document}